\documentclass{taira}
\usepackage{color}
\usepackage{tikz}
\usetikzlibrary{shapes}
\usepackage{amssymb}
\usepackage{amsmath}
\usepackage{xcolor}
\usepackage{graphicx}

\begin{document}
\definecolor{c1}{rgb}{0,114,189}
\definecolor{c2}{rgb}{237,177,32}
\definecolor{c3}{rgb}{217,83,25}
\definecolor{c4}{rgb}{126,47,142}
\newcommand{\markerone}{\raisebox{0.0pt}{\tikz{\node[scale=0.7,regular polygon, circle, fill={rgb,255:red,0; green,114; blue,189}](){};}}}
\newcommand{\markertwo}{\raisebox{0.0pt}{\tikz{\node[scale=0.5, regular polygon, regular polygon sides=3, fill={rgb,255:red,237; green,177; blue,32}](){};}}}
\newcommand{\markerthree}{\raisebox{0.0pt}{\tikz{\node[scale=0.5,regular polygon, regular polygon sides=3,fill={rgb,255:red,217; green,83; blue,25},rotate=-90](){};}}}
\newcommand{\markerfour}{\raisebox{0.0pt}{\tikz{\node[scale=0.5,regular polygon, regular polygon sides=3,fill={rgb,255:red,126; green,47; blue,142},rotate=-180](){};}}}

\newtheorem{lemma}{Lemma}
\newtheorem{corollary}{Corollary}

\shorttitle{Three-dimensional flows over finite-AR wings under tip effects} 

\shortauthor{K. Zhang et al.} 

\title{On the formation of three-dimensional flows over finite-aspect-ratio wings under tip effects} 

\author
 {
 Kai Zhang\aff{1,2}
  \corresp{\email{kzhang3@ucla.edu}},
  Shelby Hayostek\aff{3},
  Michael Amitay\aff{3},
  Wei He\aff{4},
  Vassilios Theofilis\aff{4,5}, 
  \and
  Kunihiko Taira\aff{1,2}
  }

\affiliation
{
\aff{1}
Department of Mechanical and Aerospace Engineering, University of California, Los Angeles, CA 90095, USA

\aff{2}
Department of Mechanical Engineering, Florida State University, Tallahassee, FL 32310, USA

\aff{3}
Department of Mechanical, Aeronautical, and Nuclear Engineering, Rensselaer Polytechnic Institute, Troy, NY 12180, USA

\aff{4}
Department of Mechanical, Materials and Aerospace Engineering, University of Liverpool, Brownlow Hill, England L69 3GH, United Kingdom

\aff{5}
Escola Politecnica, Universidade S\~ao Paulo, Avda. Prof. Mello Moraes 2231,  CEP 5508-900, S\~ao Paulo-SP, Brasil 
}
\maketitle

\begin{abstract}
We perform direct numerical simulations of flows over finite-aspect-ratio unswept NACA 0015 wings at $Re=400$ over a range of angles of attack (from $0^{\circ}$ to $30^{\circ}$) and aspect ratios (from 1 to 6) to characterize the tip effects on separation and wake dynamics. This study focuses on the development of three-dimensional separated flow over the wing, and discuss flow structures formed on the wing surface as well as in the far field wake. Vorticity is introduced from the wing surface into the flow in a predominantly two-dimensional manner. The vortex sheet from the wing tip rolls up around the free end to form the tip vortex. At its inception, the tip vortex is weak and its effect is spatially confined.  As the flow around the tip separates, the tip effects extend farther in the spanwise direction, generating noticeable three dimensionality in the wake. For low-aspect-ratio wings ($AR\approx 1$), the wake remains stable due to the strong tip-vortex induced downwash over the entire span. Increasing the aspect ratio allows unsteady vortical flow to emerge away from the tip at sufficiently high angles of attack.  These unsteady vortices shed and form closed vortical loops.  For higher-aspect-ratio wings ($AR\gtrsim 4$), the tip effects retard the near-tip shedding process, which desynchronizes from the two-dimensional shedding over the mid-span region, yielding vortex dislocation.  At high angles of attack, the tip vortex exhibits noticeable undulations due to the strong interaction with the unsteady shedding vortices. The spanwise distribution of force coefficients is found to be related to the three-dimensional wake dynamics and the tip effects. Vortical elements in the wake that are responsible for the generation of lift and drag forces are identified through the force element analysis. We note that at high angles of attack, a stationary vortical structure forms at the leading edge near the tip, giving rise to locally high lift and drag forces.  The analysis performed in this paper reveals how the vortical flow around the tip influences the separation physics, the global wake dynamics, and the spanwise force distributions.
\end{abstract}

\section{Introduction}
\label{sec:intro}
Separated flows over lifting surfaces have been studied extensively due to their critical importance in aerodynamics and hydrodynamics. Substantial researches have been dedicated to the understanding of separated flows over canonical airfoils to reveal the separation mechanism and their influences on the aerodynamic characteristics \citep{anderson2010fundamentals}. In many of the fundamental studies, the assumption of two-dimensional (or quasi-two-dimensional) flow was applied to assess the performance of airfoils, offering valuable insights \citep{abbott1959theory,pauley1990structure, huang2001surface, yarusevych2009vortex, he2017linear, rossi2018multiple}.

Over the past couple of decades, there have been a large number of studies focusing on three-dimensional post-stall flows at angles of attack much higher than what were traditionally considered.  These studies were performed to examine fundamental aspects of flows related to biological fliers and swimmers as well as small-scale aircraft that experience large-amplitude perturbations in flight \citep{ellington1996leading, dickinson1999wing, eldredge2019leading}.  The lifting surfaces for these flows have some uniqueness.  The wings have low-aspect-ratio planforms and generate prominent tip vortices at high angles of attack \citep{taira2009three, lentink2009rotational}.  For these reasons, the resulting wakes are highly three-dimensional with complex nonlinear dynamics generated by the interactions among the wake vortices.   Under the tip effects, the wake dynamics of finite wings can be significantly different from its two-dimensional analogue. The three-dimensional nature of the wakes requires global analysis of the separation dynamics.

Separated flows over finite-aspect-ratio wings in pure translation have been examined with surface oil visualizations  \citep{gregory1971progress,wang1976separation,winkelmann1980effects}. From these visualizations, \citet{winkelman1980flowfield} proposed a flow field model for a post-stall rectangular wing. This model consists of a pair of counter-rotating tip vortices at the two ends of the wing and the mushroom-like three-dimensional separation bubble.
Later, by smoke visualization, \citet{freymuth1987further} revealed the intricate three-dimensional wake structures behind finite wings.  On the aerodynamic characterizations of low-aspect-ratio wings at low Reynolds numbers, \citet{pelletier2000low}, \citet{torres2004low} and \citet{ananda2015measured} obtained a wealth of force and pitching moment data for a large collection of aspect ratios, angles of attack, and planform geometries.

Direct numerical simulations have provided detailed insights into separated flows over finite-aspect-ratio wings in translation. The wake structures behind an impulsively started plate were analyzed by \citet{taira2009three} using direct numerical simulation at Reynolds numbers of 300 and 500. In the initial stage of wake development, the wake vortices share the same structures for all aspect ratios. At large time, the tip vortices significantly influence the vortex dynamics and the corresponding forces on the wings. The long-term stability of the wake was found to be dependent on the angle of attack, aspect ratio and the Reynolds number.
\citet{lee2012vorticity} applied the force element theory \citep{chang1992potential} to the impulsively started flows around low-aspect-ratio wings and identified the wake regions responsible for the force generation. They have shown that the tip vortices exert lift force during the start-up maneuver.  More recently, \citet{garmann2017investigation} conducted high-fidelity large eddy simulations of flow over finite-aspect-ratio wings at $\Rey=2\times10^5$. Intricate details of the vortical structures near the wing tip were revealed. 

Stability analysis have also been carried out to study the instabilities in the wake of finite-aspect-ratio wings. \citet{he2017linear_b} conducted linear stability analysis on the flows over finite elliptical wings. Two classes of linearly unstable perturbations were identified, namely the highly-amplified symmetric modes and the weakly-amplified antisymmetric disturbances.  \citet{brion2019transient} performed transient growth analysis on the wake of finite NACA 0012 wings. The optimal linear perturbation was found to be located near the surface of the wing in the form of chord-wise periodic structures whose strength decreases from the root towards the tip. \citet{edstrand2018parallel} carried out parabolized stability analysis of the tip vortex generated from a finite NACA 0012 wing at the Reynolds number of 1000. The insights obtained from their analysis have enabled effective design of active control techniques for the attenuation of the tip vortex \citep{edstrand2018active}. 

We note that three-dimensional separated flows over finite-aspect-ratio wings have also been extensively studied for wings under unsteady maneuvers, including surging \citep{chen2010leading,mancini2015unsteady}, rotation \citep{lentink2009rotational,harbig2013reynolds,jones2016characterizing}, pitching \citep{buchholz2006evolution, green2008effects, yilmaz2012flow,jantzen2014vortex}, heaving \citep{visbal2013three}, plunging \citep{akkala2017vorticity}, and flapping \citep{birch2001spanwise,dong2006wake,shyy2010recent}.  These works have closely examined the formation of large-scale vortical structures, with particular emphasis on the leading-edge vortex and its influence on added lift \citep{eldredge2019leading}. 

Despite the vast literature, there are still fundamental questions left unanswered concerning three-dimensional separated flows over finite-aspect-ratio wings in pure translation. It is generally known that the three-dimensional wake dynamics described in the previous works are closely related with the tip effects. However, the detailed process of how the tip-imposed three dimensionality is introduced into the flow remains elusive. The formation of the wake vortical structures under the tip effects, as well as the dynamical relations between those structures requires further analysis.  

To address the aforementioned questions, we carry out a large number of direct numerical simulations of flows over finite-aspect-ratio NACA 0015 wings.  By focusing on the development of the three-dimensional wake, we characterize the tip effects on separation and wake dynamics.  In what follows, we present the computational set-up and its validation in \S \ref{sec:setup}. The results are discussed in \S \ref{sec:results} starting with the flow physics on the wing surface in \S \ref{sec:fluxAndSurface}. A detailed look at the three-dimensional wake structures is provided in \S \ref{sec:wakeTopology}, followed by the spectral analysis of the wake in \S \ref{sec:wakeSpectral}. A classification of the wake states for cases with different angles of attack and aspect ratios is provided \S \ref{sec:classification} The aerodynamic forces and their relationship with the wake dynamics are presented in \S \ref{sec:forces}. We conclude this study by summarizing our findings in \S \ref{sec:conclusion}.

\section{Computational set-up}
\label{sec:setup}
\subsection{Numerical approach}
\begin{figure}
	\centering
	\includegraphics[scale=0.55]{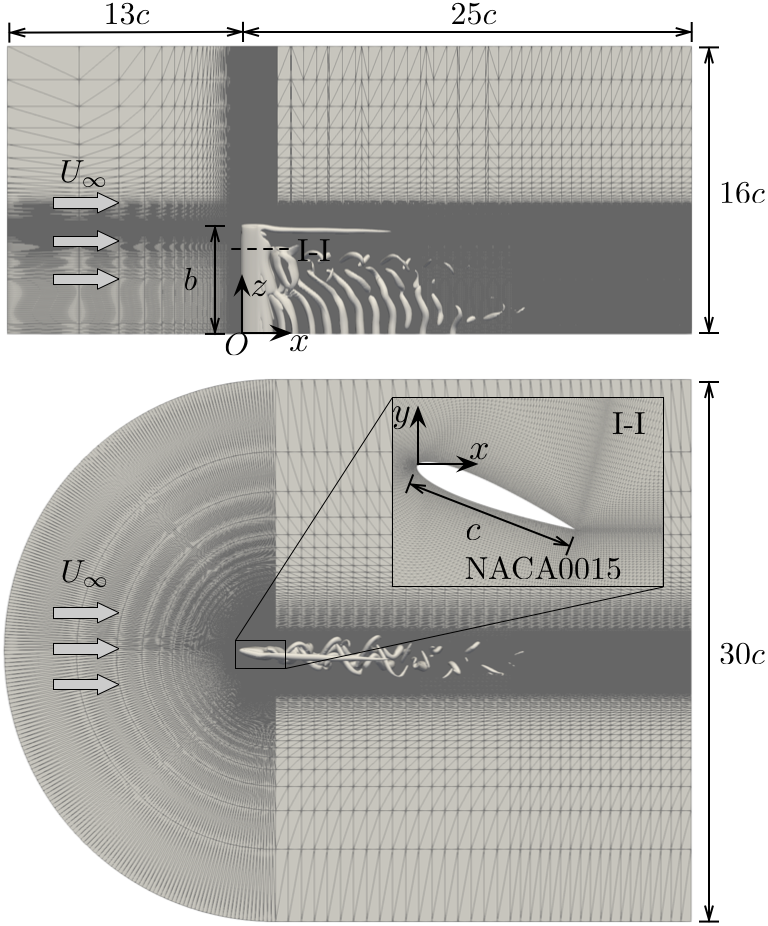}
	\caption{
	The computational setup for the case of $(\alpha,AR)=(22^{\circ},6)$}
	\label{fig:MeshCoord}
\end{figure}

We numerically analyze incompressible flow over a NACA 0015 wing with a straight cut tip.  The wing is situated in uniform flow with velocity $U_\infty$ as shown in figure \ref{fig:MeshCoord}.  The three-dimensional flow over the finite wing is studied by numerically solving the Navier--Stokes equations
\begin{subeqnarray}
 \frac{\partial \boldsymbol{u}}{\partial t} +  \boldsymbol{u} \cdot \boldsymbol{\nabla} \boldsymbol{u} & = & -\boldsymbol{\nabla} p +\displaystyle{\frac{1}{\Rey}} \boldsymbol{\nabla}^2 \boldsymbol{u},\\
 \boldsymbol{\nabla} \cdot \boldsymbol{u} & = & 0,
\label{equ:NS}
\end{subeqnarray}
where $\boldsymbol{u}=(u_x,u_y,u_z)$ is the velocity vector and $p$ is the pressure. An incompressible flow solver \emph{Cliff} (in \emph{CharLES} software package, Cascade Technologies, Inc.) based on the finite-volume formulation with second-order accuracy in time and space \citep{ham2004energy,ham2006accurate} is used in the present study. 

For non-dimensionalization, we normalize the spatial variables by the chord length $c$, velocities by the freestream velocity $U_{\infty}$, and time by $c/U_{\infty}$.  The Reynolds number in the momentum equation (\ref{equ:NS}) is defined as $\Rey \equiv U_{\infty}c/\nu$, where $\nu$ is the kinematic viscosity.  In this work, we set the Reynolds number to $\Rey=400$ so that the flow remains laminar.  As we are interested in the three-dimensional flow physics imposed by the tip effects, we simulate the half-wing model by prescribing the symmetry boundary condition along the mid-span.  Denoting the length of the half wing as $b$, the (half-wing) aspect ratio is defined as $AR=b/c$ and is varied from 1 to 6. The angle of attack $\alpha$ is varied from from $0^{\circ}$ to $30^{\circ}$. 

The computational domain covers $(x,y,z)\in[-13,25]\times[-15,15]\times[0,16]$, where $x$, $y$ and $z$ are the streamwise, crossflow and spanwise coordinates. This results in a maximum blockage ratio of 0.8\% for the case of $(\alpha,AR)=(30^{\circ},6)$. The origin of the Cartesian coordinate system is placed at the leading edge of the wing on the mid-span plane ($z=0$) plane. The computational domain is discretized with a C-type grid with the mesh refined in the vicinity of the wing as well as its wake. The adequacy of the grid resolution is examined in \S \ref{sec:VV}. 

The inlet and far-field boundaries are prescribed with freestream $\boldsymbol{u}=(U_{\infty}, 0, 0)$. The convective boundary condition ($\partial \boldsymbol{u}/\partial t+U_{\infty}\partial \boldsymbol{u}/\partial x=0$) is applied at the outlet to allow wake structures to leave the domain without disturbing the near-field flow. The wing surface is treated with the no-slip boundary condition. The simulations are started with uniform flow, and statistics are recorded only after the initial transients are flushed out of the computational domain (typically 25 convective time units). Flow statistics are collected with over 100 convective time units to ensure statistical convergence. 

\subsection{Verification and validation}
\label{sec:VV}
\begin{table}
 \begin{center}
  \begin{tabular}{lccccc}
    Mesh  & $N_p$   &   $N_s$ & $N_z$ & $N_{CV}$ & $\mathrm{CFL}$   \\[3pt]
       medium   & 60 & 80 & 120 & $14.8\times 10^6$ & 1.0\\
       refined  & 80 & 100 & 150 & $20.6\times 10^6$ & 0.5\\
  \end{tabular}
  \caption{
  Setups for the different meshes for the case of $(\alpha,AR)=(22^{\circ}, 6)$. $N_p$ and $N_s$ are the numbers of grid points on the pressure and suction side of the airfoil. $N_z$ is the number of grids along the wing span. $N_{CV}$ is the total number of control volumes, and $\mathrm{CFL}$ is the Courant number used in the simulations.}
 \end{center}
 \label{table:mesh}
\end{table}

\begin{figure}
    \centering
    \includegraphics[scale=0.47]{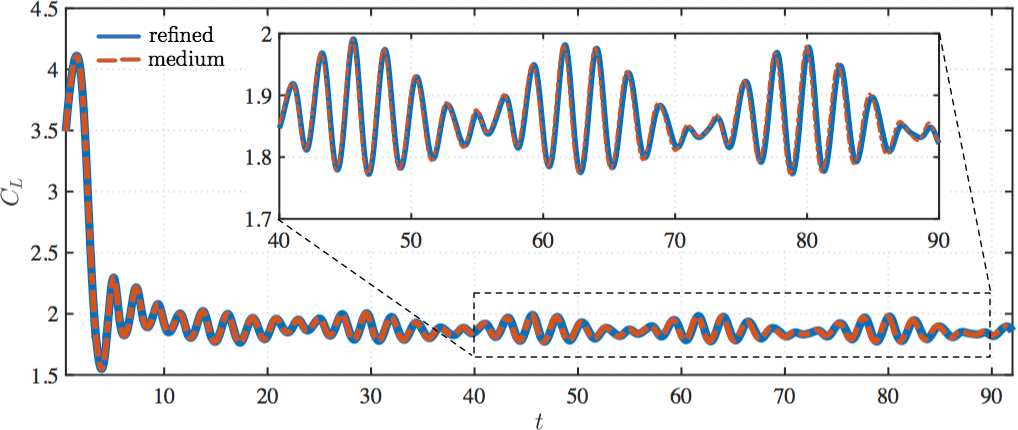}
    \caption{
    Time histories of lift coefficient obtained from two sets of meshes for the case of $(\alpha, AR)=(22^{\circ}, 6)$.}
    \label{fig:meshTest}
\end{figure}

To verify the convergence of the numerical results with respect to the grid resolution, we examine the case of $(\alpha,AR)=(22^{\circ}, 6)$ with two sets of meshes described in table 1.  
For the refined mesh, the temporal resolution is also increased by restricting the maximum CFL number to be half of the medium mesh.  We compare the forces on the wing evaluated from the two meshes to assess the mesh requirement.  
The time history of the lift force obtained from these meshes are reported in figure \ref{fig:meshTest}.  Throughout this study, the lift and drag forces are reported in their non-dimensional forms through
\begin{subeqnarray}
    C_L=\displaystyle{\frac{\int_{S}(-p\boldsymbol{n}+\boldsymbol{\tau}) 
    \cdot\boldsymbol{e_y} \mathrm{d}S}{\frac{1}{2}U_{\infty}^2 bc}} 
    \quad 
    \text{and}
    \quad
    C_D=\displaystyle{\frac{\int_{S}(-p\boldsymbol{n}+\boldsymbol{\tau})
    \cdot\boldsymbol{e_x}\mathrm{d}S}{\frac{1}{2}U_{\infty}^2 bc}},
\label{equ:force}
\end{subeqnarray}
where $S$ denotes the wing surface, $\boldsymbol{n}$ is the unit normal vector pointing outward from the wing surface. Unit vectors $\boldsymbol{e_y}$ and $\boldsymbol{e_x}$ are in the lift and drag directions. Moreover, $\boldsymbol{\tau}=\mu \boldsymbol{\omega}\times \boldsymbol{n}$ is the skin-friction vector ($\mu$ is the dynamic viscosity of fluid and $\boldsymbol{\omega}=\boldsymbol{\nabla}\times \boldsymbol{u}$ is the vorticity vector). 
Once the initial transient settles (from the use of uniform flow as the initial condition), the lift coefficients exhibit quasi-periodic oscillations with low-frequency beating. This is resolved well with both mesh resolutions even at large times. Thus, the medium resolution grid is deemed adequate to yield accurate results and is used throughout this study.

\begin{figure}
\centering
\includegraphics[scale=0.47]{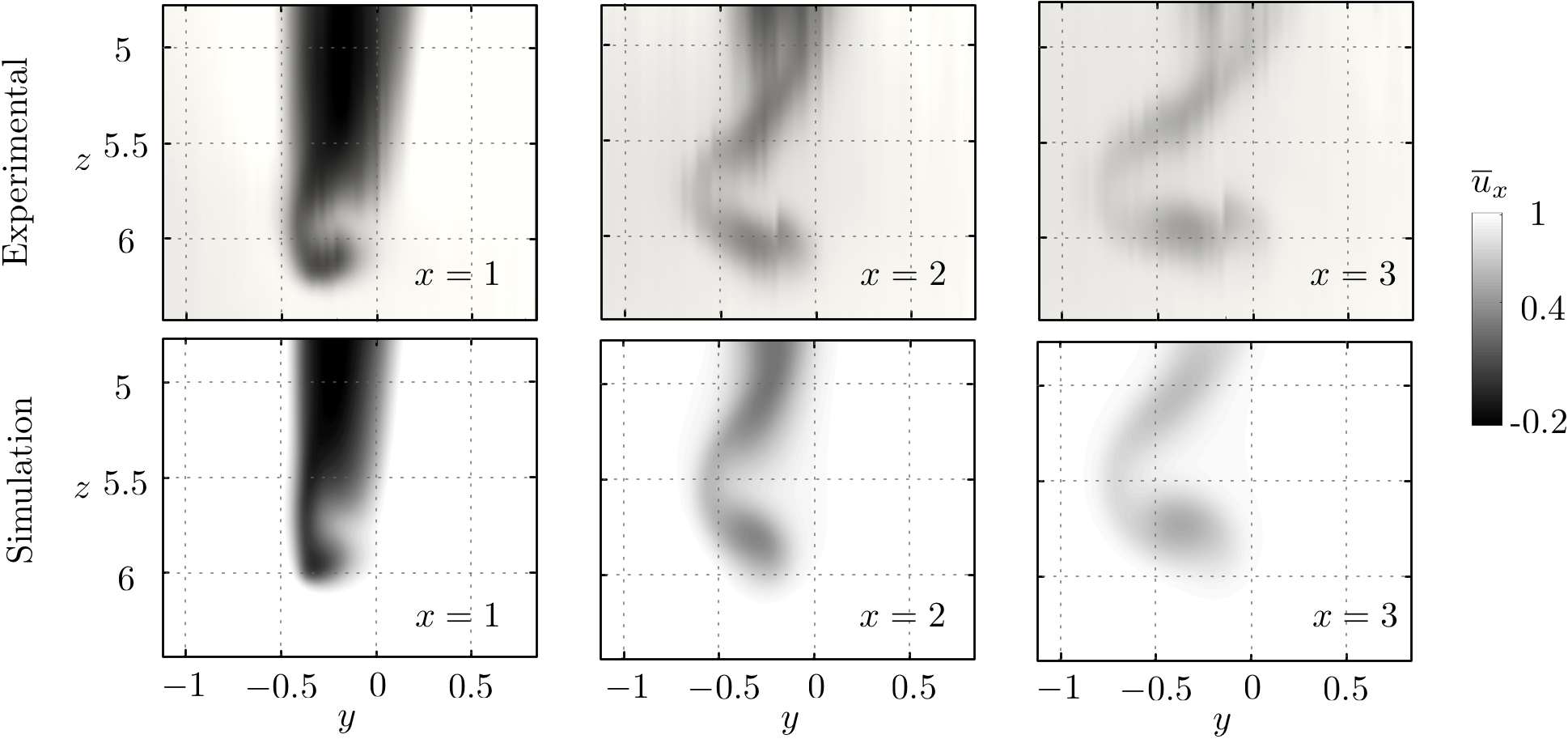}
\caption{Comparison of time-averaged streamwise velocity $\overline{u}_x$ between the experiment and the present simulation at selected streamwise locations aft of a wing of $(\alpha,Re)=(22^{\circ},6)$. The experiments are conducted at $\Rey=600$ in the water tunnel at Rensselaer Polytechnic Institute, while the numerical simulation is performed at $\Rey=400$. }
\label{fig:expVal}
\end{figure}

To validate our computational setup, we compare the numerical results against the stereoscopic particle image velocimetry (SPIV) measurements from water tunnel experiments conducted at the Rensselaer Polytechnic Institute. Detailed descriptions of the experimental setup are documented in \citet{hayostek2018three}. 
In contrast to the symmetry boundary condition used for the mid-span in the present simulations, a wall is present at the root plane in the water tunnel. For this reason, comparison between the numerical and experimental results is focused near the tip region for the case of $(\alpha,AR)=(22^{\circ}, 6)$, where the influence of the root boundary condition is small.
The time-averaged streamwise velocity fields are shown in figure \ref{fig:expVal} for three streamwise locations at $x=1$, $2$ and $3$. While the water tunnel experiments are conducted at a larger Reynolds number of 600, the flow remains laminar and is in good agreement with that from the present numerical simulation, confirming the fidelity of the numerical results. 

\begin{figure}
\centering
\includegraphics[scale=0.47]{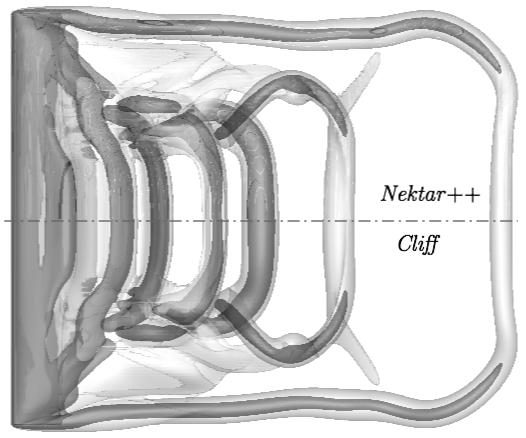}
\caption{Comparison of the vortical structures obtained from \textit{Nektar++} (top) and \textit{Cliff} (bottom) at $t=9.25$ for the case $(\alpha,AR)=(22^{\circ},4)$. Both simulations are started with uniform initial condition. Shown are iso-surfaces of $\|\boldsymbol{\omega}\|=1$ in transparent gray and $Q=1$ in dark gray. }
\label{fig:nektarVal}
\end{figure}

We further validate the present numerical results against those from another flow solver \textit{Nektar++} that is based on the spectral/hp element method \citep{cantwell2015nektar}. The computation is carried out with an independent numerical setup \citep{he2019wake}. We compare the wake vortical structures obtained by \textit{Nektar++} (top) and \textit{Cliff} (bottom) for the same case of $(\alpha,AR)=({22^{\circ},4})$ at $t=9.25$ in figure \ref{fig:nektarVal}. Shown are the iso-surfaces of $\|\boldsymbol{\omega}\|=1$ (vorticity magnitude) and $Q = 1$ (second invariant of velocity gradient tensor). The vortical structures obtained from the two solvers agree in an excellent manner with each other even for the smallest flow details. Based on the above verification and validation, we have ensured that the current computational setup is reliable. We now proceed to present our results in the next section.

\section{Results}
\label{sec:results}

\begin{figure}
    \centering
    \includegraphics[scale=0.45]{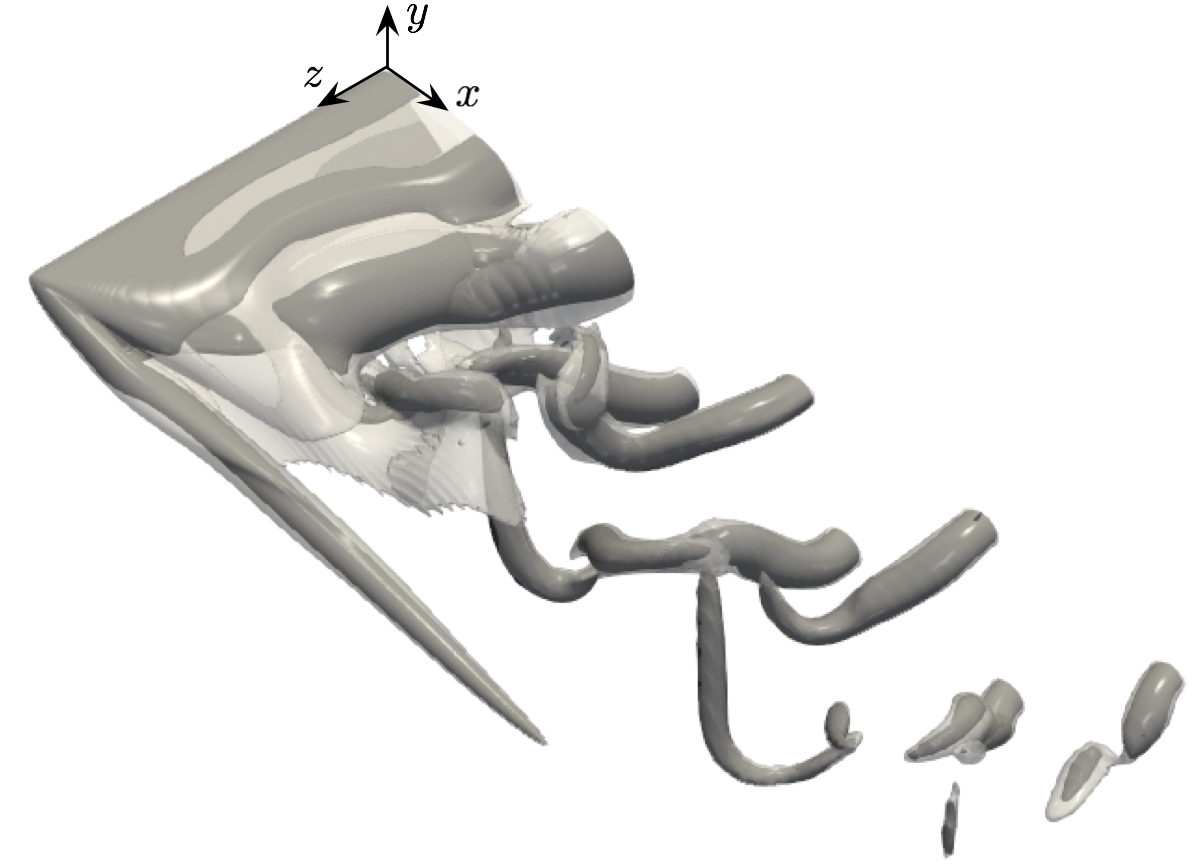}
    \caption{
    Perspective view the vortical structures for the case of $(\alpha, AR)=(22^{\circ},4)$. Iso-surfaces of $Q=1$ are shown in dark gray to visualize vortex cores and $\|\omega\|=2$ in transparent gray to represent the vorticity sheet.}
    \label{fig:Wake_Q}
\end{figure}

The wake behind a finite-aspect-ratio wing exhibits rich three-dimensional flow features due to the tip effects, as shown in figure \ref{fig:Wake_Q}. In this section, we begin the discussions by focusing on the wing surface to look for the source of vorticity. Next, we describe in detail the three-dimensional wake dynamics.  The aerodynamic forces are also discussed with an emphasis on their relationship with the three-dimensional wake structures.

\subsection{Boundary vorticity flux and skin-friction lines }
\label{sec:fluxAndSurface}

We begin our analysis by examining the introduction of vorticity to the flow from the wing surface. In incompressible flow, vorticity is only generated on the solid surface and diffuses into the flow \citep{morton1984generation,hornung1989vorticity,wu2007vorticity}. The rate at which vorticity is created at the surface is given by the wall-normal boundary vorticity flux 
\begin{equation}
	\boldsymbol{\sigma}=-\nu\boldsymbol{\Sigma}\boldsymbol{n},
\label{equ:vorticityFlux}
\end{equation}
where $\boldsymbol{\Sigma}=(\boldsymbol{\nabla}\boldsymbol{\omega})_0$ is the vorticity gradient tensor at the surface. 

\begin{figure}
	\centering
	\includegraphics[scale=0.46]{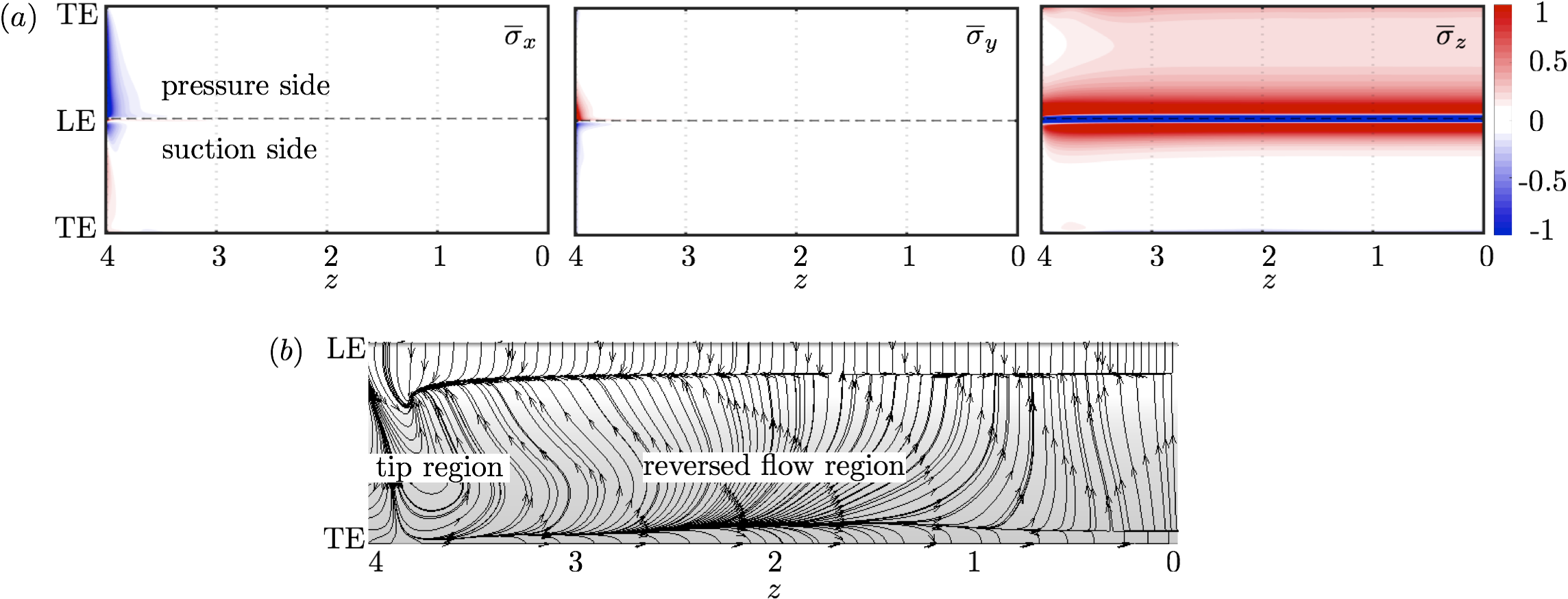}
	\caption{
	Time-averaged flow behavior at the wing surface for $(\alpha, AR)=(22^{\circ}, 4)$. (\textit{a}) Three components of the wall-normal boundary vorticity flux. The surface of the wing is unfolded to a plane by cutting along the trailing edge (TE). Dashed line indicates the leading edge (LE) of the wing. (\textit{b}) Skin-friction lines on the suction surface.
	}
\label{fig:flux}
\end{figure}

Let us show the vorticity flux along the wing surface for a representative case of $(\alpha,AR)=(22^{\circ},4)$. The time-averaged wall-normal boundary vorticity flux components $\overline{\sigma}_x$, $\overline{\sigma}_y$ and $\overline{\sigma}_z$ are plotted in figure \ref{fig:flux}(\textit{a}) to identify the local generation of vorticity.  As expected, the spanwise vorticity flux $\overline{\sigma}_z$ dominates over the surface of the wing.
Negative $\overline{\sigma}_z$ is contained within a narrow region along the leading edge of the wing. This region injects a large amount of negative $\omega_z$ into the flow, forming the leading-edge vortex sheet. Compared to $\overline{\sigma}_z$, the other two components of the vorticity flux appear negligible for the majority of the span. It is only at the close vicinity of the wing tip that considerable level of $\overline{\sigma}_x$ and $\overline{\sigma}_y$ can be found. The creation of vorticity on the finite-aspect-ratio wing is not strongly affected by the tip effects and is predominantly two-dimensional along the span.

Once the vorticity is generated at the wing surface, it is diffused into the flow and is convected by the freestream. The developed flow imprints the skin-friction field $\boldsymbol{\tau}$ on the wall, as shown in figure \ref{fig:flux}(\textit{b}). Three regions can be recognized from the skin-friction field. The region close to the leading edge away from the tip is occupied by the nearly two-dimensional boundary layer, which separates from the wing in a uniform manner. 
Post separation, a region of reversed flow appears on the suction side of the wing where the skin-friction lines exhibit spanwise component of the flow. The dominant three dimensionality is observed over the wing-tip region with a three-dimensional swirl pattern, as reported in the experimental work of \citet{winkelman1980flowfield}. 

\begin{figure}
\centering
\includegraphics[scale=0.47]{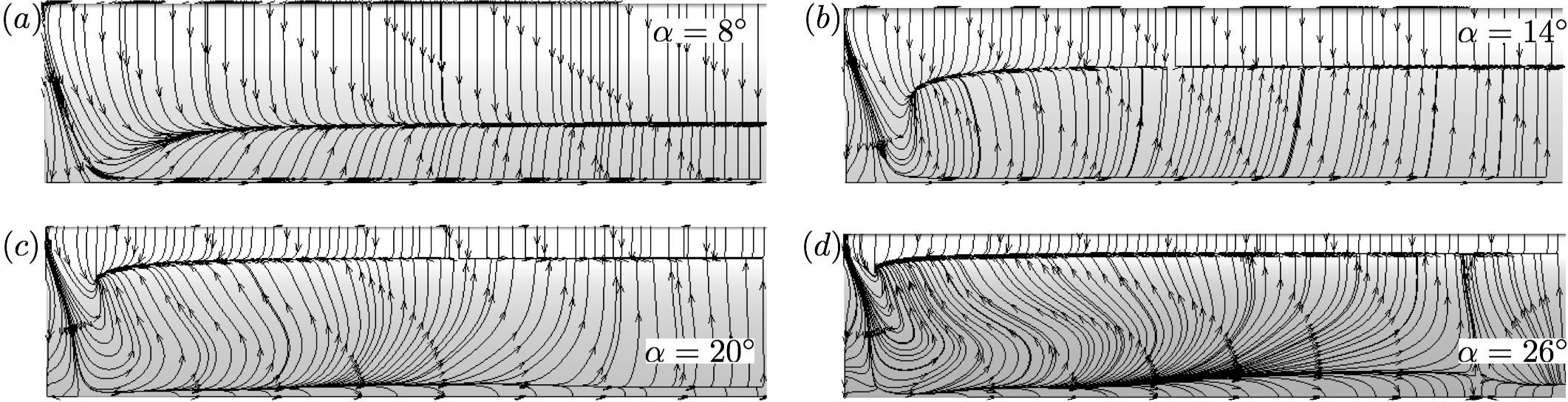}
\caption{
Time-averaged skin-friction line patterns on the suction side for the cases of $AR=4$ for $(a)$ $\alpha=8^{\circ}$, $(b)$ $\alpha=14^{\circ}$, $(c)$ $\alpha=20^{\circ}$ and $(d)$ $\alpha=26^{\circ}$.}
\label{fig:surfaceLines_AoA}
\end{figure}

To assess the influence of the angle of attack on the near-surface flow pattern, we visualize the time-averaged skin-friction lines for the cases of $AR=4$ with different $\alpha$ in figure \ref{fig:surfaceLines_AoA}. For higher angles of attack, the flow separation occurs more upstream, and the spanwise extent of the boundary layer that is affected by the tip effects becomes smaller. 
The skin-friction lines in the recirculation region becomes more three-dimensional at higher $\alpha$.

The development of the three-dimensional flow can be envisioned from the boundary vorticity flux and skin-friction lines. 
Vorticity is introduced from the wing surface in a predominantly two-dimensional manner. The vortex sheets emanate from the leading edge and the wing tip. While the roll up of the leading-edge vortex sheet is two-dimensional, the roll-up from the tip is fundamentally three-dimensional yielding a streamwise tip vortex. As the tip vortex develops, its influence grows from being locally-confined to imposing spanwise variations, as evident in figure \ref{fig:Wake_Q}. We discuss the three-dimensional wake dynamics in the following sections. 

\subsection{Three-dimensional wake structures}
\label{sec:wakeTopology}
\subsubsection{An overview of the vortical structures}
\label{sec:keyFeatures}

\begin{figure}
\centering
\includegraphics[scale=0.52]{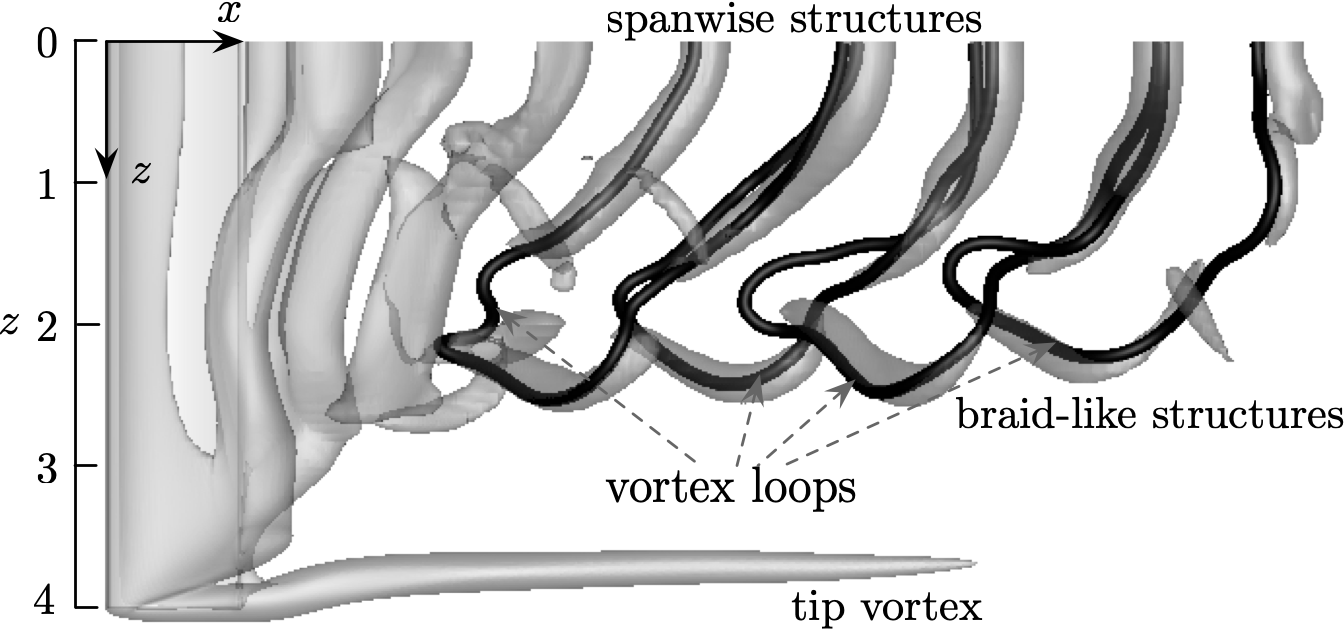}
\caption{
Top view of the wake for the case of $(\alpha,AR)=(22^{\circ},4)$. The black lines represents the vorticity lines. Iso-surfaces of $Q=1$ are shown in transparent gray.}
\label{fig:generalDescription}
\end{figure}

Let us continue to examine the three-dimensional wake structures for the representative case of $(\alpha, AR) = (22^{\circ}, 4)$. A top view of the wake vortical structures is shown in figure \ref{fig:generalDescription}. 
As also seen in figure \ref{fig:Wake_Q}, the wake can be divided into the tip vortex region at the free end and the unsteady vortex shedding region near the mid-span. 
The tip vortex appears practically steady over time in the current case. The unsteadiness in the tip vortex becomes noticeable for cases with smaller $AR$ and larger $\alpha$, as discussed later in \S \ref{sec:AoAAREffect} and \S \ref{sec:tipVortex}.
The unsteady region of the wake can be further divided into two subregions.
Adjacent to the mid-span ($z\lesssim 1.5$), the wake is dominated by vortex cores that are aligned predominantly with the spanwise direction. These vortices are formed by the roll-up of the vortex sheets from the leading and trailing edges, and shed alternatively into the wake. 
On the other hand, the shedding vortices at $2\lesssim z\lesssim 2.5$ feature braid-like structures comprised of both the streamwise vorticity $\omega_x$ and crossflow vorticity $\omega_y$. These vortical structures shed at the same frequency with the spanwise vortices. 

The complex vortical structures in the unsteady region can be better understood by visualizing the vorticity lines, as depicted by the black solid lines in figure \ref{fig:generalDescription}. Each vortex core in the braid-like region connects with a pair of counter-rotating spanwise vortical structures.  These structures together form a closed vortex loop (considering the symmetry boundary condition at the mid-span). The unsteady wake region is constituted by a series of such vortex loops arranged in a zig-zag manner. It is thus clear that the braid-like structures play the role of closing the spanwise vortex system. This is in accordance with Helmholtz's second law which states that a vortex filament can not terminate in a fluid \citep{batchelor1967introduction}. Similar vortex line models have also been proposed to describe the wakes of finite-span circular cylinders \citep{taneda1952studies,levold2012viscous}.

\subsubsection{Effects of $AR$ and $\alpha$ on the wake}
\label{sec:AoAAREffect}

\begin{figure}
	\centering
	\includegraphics[scale=0.499]{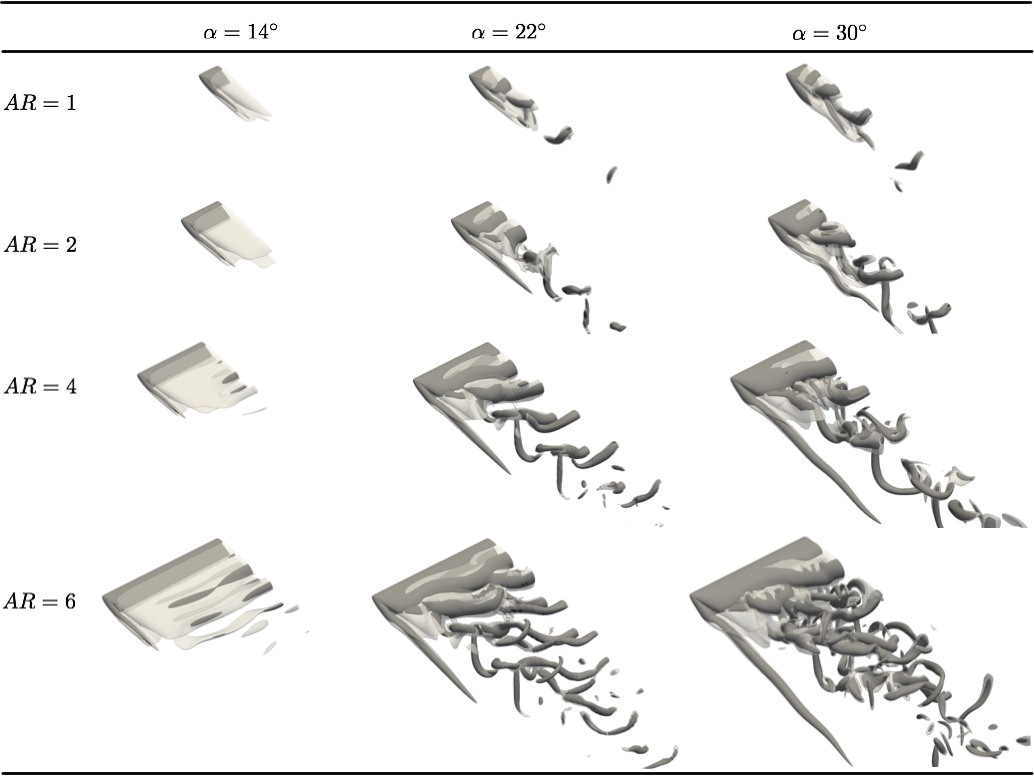}
	\caption{
	Perspective views of the wake vortices behind the finite wings with $AR=1, 2, 4$ and 6, $\alpha=14^{\circ}, 22^{\circ}$ and $30^{\circ}$. Vortical structures are visualized by the iso-surfaces of $\|\boldsymbol{\omega}\|=2$ in light gray and $Q=1$ in dark gray.}
\label{fig:ARAoAEffect}
\end{figure}

Separated flows behind finite-aspect ratio wings are strongly influenced by the aspect ratio $AR$ and the angle of attack $\alpha$.  Let us show some representative wakes for a range of aspect ratios and angles of attack in figure \ref{fig:ARAoAEffect}.  The short wings with $AR=1$ generates different wake features compared with those discussed in \S \ref{sec:keyFeatures}. Stable three-dimensional flow is observed at $\alpha=14^{\circ}$, while the analogous two-dimensional flow at the same angle of attack exhibits periodic unsteady shedding.  This indicates again that the tip effects stabilize the local flow, which in the current case of $AR=1$ is the entire span. As the angle of attack is increased to $22^{\circ}$ and $30^{\circ}$, the wake instability becomes sufficiently strong to overcome the stabilizing effect from the tip, resulting in periodic shedding of the wake vortices. Originated from the leading edge with predominately spanwise vorticity, the shedding vortices on the suction side gradually morph into the hairpin vortices with their legs extending far upstream. The tip vortex cannot be clearly observed due to its strong interaction with the unsteady shedding vortices. Similar vortical features have also been reported in the wake of a flat-plate wing at similar Reynolds numbers by \citet{taira2009three}.

For $AR=2$, the increased spanwise extent alleviates the strong interaction between the tip vortex and unsteady shedding vortices as the case in $AR=1$ wing. This enables the tip vortex to remain relatively steady despite the unsteadiness presented in the rest of the flow. The wake at $\alpha=14^{\circ}$ exhibits weak periodic shedding according to a close examination of the lift force, although it may not be clearly visible in figure \ref{fig:ARAoAEffect}. At $\alpha=22^{\circ}$, the unsteady shedding vortices mainly take the form of the braid-like structures, without clear spanwise vortex cores. As the angle of attack further increases to $\alpha=30^{\circ}$, the unsteady vortex shedding grows stronger. Under its effect, the tip vortex exhibits noticeable undulations along the streamwise direction.

Increasing the aspect ratio to $AR=4$ reveals the spanwise vortex shedding region. The spanwise vortical structures are quite weak at $\alpha=14^{\circ}$ and appear only in the direct vicinity of the mid-span. With the increase in angle of attack, the unsteady shedding vortices grow stronger and extend farther into the wake. The tip vortex elongates compared with that at lower angles of attack. At $\alpha=30^{\circ}$, small-scale streamwise vortical structures emerge in the spanwise shedding region, making the wake complex. The streamwise undulations in the tip vortex become less significant compared with that in the case of $( \alpha,AR)=(30^{\circ},2)$.

As we consider higher aspect ratio of 6, the spanwise shedding takes place over an expanded region for $\alpha=14^{\circ}$ and $22^{\circ}$. An additional feature that becomes clear at such high aspect ratio is the vortex dislocation, which divides the spanwise shedding region into two parts. This phenomenon is responsible for the low-frequency beating in the lift coefficients as shown in figure \ref{fig:meshTest}, and will be examined in detail in \S \ref{sec:wakeSpectral}. At $\alpha=30^{\circ}$, smaller-scale streamwise vortical structures reminiscent of those induced by the Floquet instability \citep{deng2017floquet,he2017linear} develop in the wake, making the unsteady shedding region even more complex.

\subsubsection{Characteristics of the tip vortex}
\label{sec:tipVortex}
\begin{figure}
	\centering
	\includegraphics[scale=0.46]{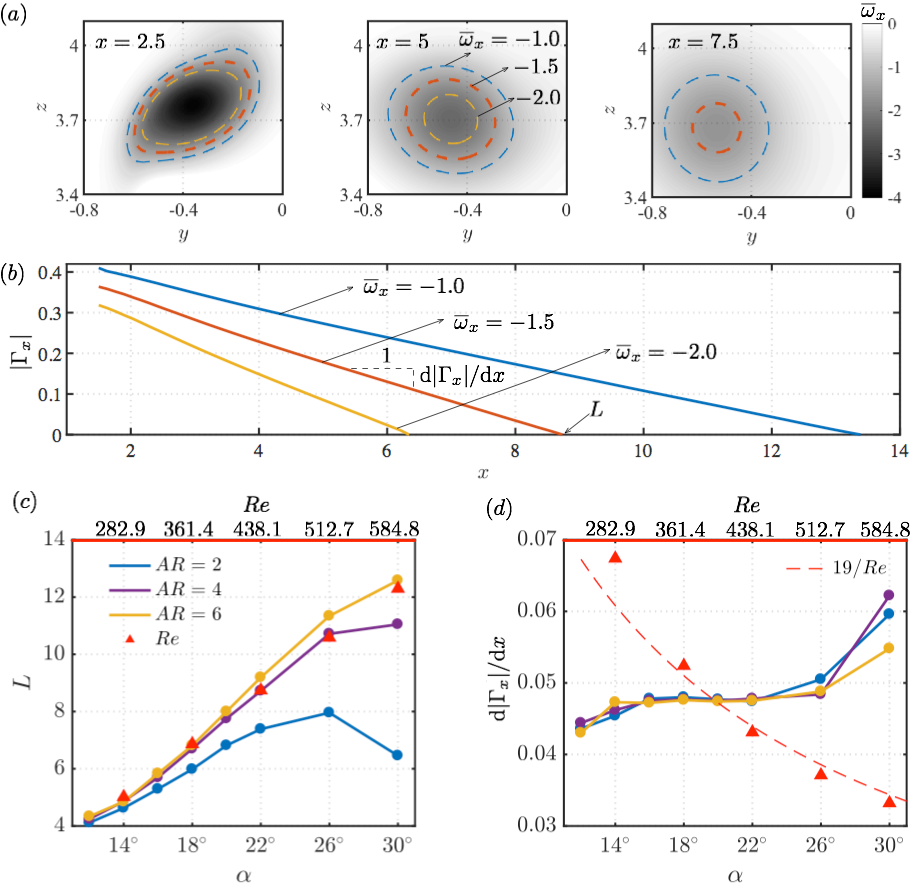}
	\caption{
	(\textit{a}) Time-averaged streamwise vorticity fields at three streamwise locations of $x=2.5$, $5$ and $7.5$ for  $(\alpha,AR)=(22^{\circ},4)$. The three dashed lines are the contour lines of $\overline{\omega}_x=-1.0$, $-1.5$ and $-2.0$. (\textit{b}) Circulations based on the three contour lines. (\textit{c}) Length $L$ of the tip vortex defined based on the zero-crossing point of $|\Gamma_x|$-$x$ line ($\Gamma_x$ evaluated based on contour of $\overline{\omega}_x=-1.5$), and (\textit{d}) decay rate $\mathrm{d}|\Gamma_x|/\mathrm{d}x$.}
\label{fig:tipVortex}
\end{figure}

We characterize the tip vortex in detail here for the representative case of $(\alpha,AR)=(22^{\circ},4)$. The time-averaged streamwise vorticity fields in the tip region are shown in figure \ref{fig:tipVortex}(\textit{a}) at $x=2.5$, 5 and 7.5. The contours of $\overline{\omega}_x=-1.0$, $-1.5$ and $-2.0$ are shown to capture the shape of the vortex core. At $x=2.5$, the three contour lines are close to each other, depicting a mango-like shape. At $x=5$, the regions enclosed by the contours of $\overline{\omega}_x=-1.5$ and $-2.0$ shrink significantly. The vorticity level inside the vortex core also decreases. Further downstream at $x=7.5$, the tip vortex core becomes more diffused, and its shape relaxes to a circle. 

The strength of the tip vortex can be quantitatively assessed through its streamwise circulation $\Gamma_x = \oint_C \boldsymbol{u}\cdot {\rm d}\boldsymbol{l}$, where the integration contour $C$ is taken as one of the three contours shown in figure \ref{fig:tipVortex}(\textit{a}). Evaluating the circulations along the streamwise direction, we obtain three $|\Gamma_x|$-$x$ profiles in figure \ref{fig:tipVortex}(\textit{b}). The strengths of the tip vortices, evaluated with the three contour levels, decay linearly in the $x$ direction. The line associated with smaller $|\overline{\omega}_x|$ measures extended length and reduced decay rate.

To characterize the strengths of the tip vortices, let us use the contour of $\overline{\omega}_x=-1.5$ and calculate the length $L$ and the decay rate $\mathrm{d}|\Gamma_x|/\mathrm{d}x$ for different cases, as summarized in figures \ref{fig:tipVortex}(\textit{c}) and (\textit{d}). We show in Appendix \ref{sec:contourLevel} that the choice of the contour level does not affect the insights drawn here. For the majority of the cases ($AR\geq 2$, $\alpha\leq 26^{\circ}$), the length of the tip vortex increases almost linearly with the angle of attack. The decay rate remains mostly constant at $\mathrm{d}|\Gamma_x|/\mathrm{d}x \approx 0.05$ due to viscous diffusion. 
Significant increase of the decay rate is observed at $\alpha\approx 26^{\circ}-30^{\circ}$ with $AR\gtrsim 2$ and $\alpha \gtrsim 20^{\circ}$ with $AR=1$. In these cases, the strong wake vortex shedding induces unsteady motion of the tip vortex, as observed in figure \ref{fig:ARAoAEffect}. This unsteadiness serves as an additional mechanism (to diffusion) to reduce the time-averaged strength of the tip vortex. As a result, the $L-\alpha$ relationship as shown in figure \ref{fig:tipVortex}(\textit{c}) no longer follows the linear growth as at smaller $\alpha$. At $AR=1$ and 2, the time-averaged tip vortex length appears to even decrease at large angles of attack.

\subsection{Spectral analysis of vortex dislocation}
\label{sec:wakeSpectral}

\begin{figure}
	\centering
	\includegraphics[scale=0.44]{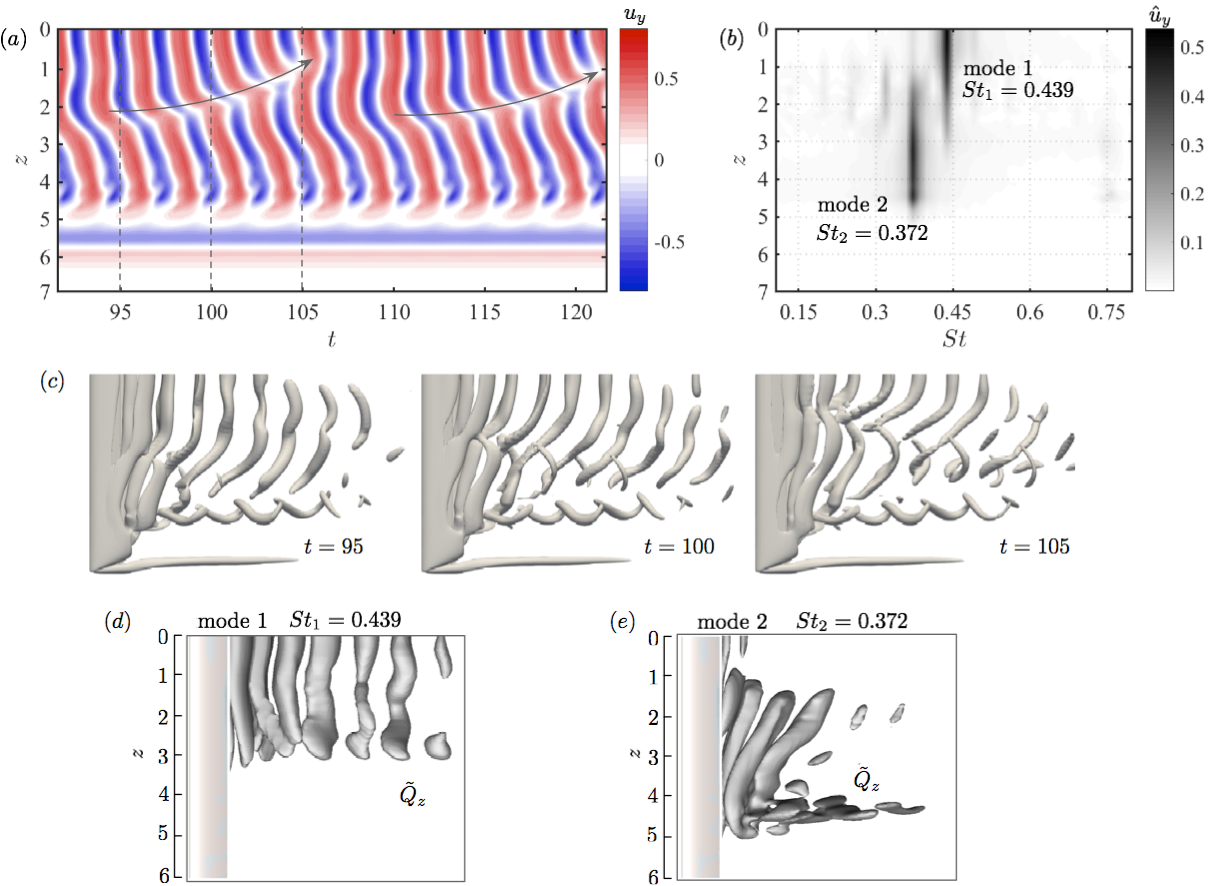}
	\caption{
	(\textit{a}) Spatial-temporal distribution of the crossflow velocity probed at $(x,y)$=$(2,-0.3)$ for the case $(\alpha,AR)=(22^{\circ},6)$. The gray arrow indicates the translating direction of the vortex dislocation. (\textit{b}) Frequency distribution of the velocity over the spanwise direction. (\textit{c}) Iso-surfaces of $Q=1$ at three time instances showing one cycle of vortex dislocation. (\textit{d}) and (\textit{e}) DMD modes ($Q$ criteria) for the two identified frequencies.}
\label{fig:DMD}
\end{figure}

As discussed in \S \ref{sec:AoAAREffect}, for large-aspect-ratio wings, vortex dislocation occurs, dividing the spanwise vortex shedding region.  Here, we focus on the particular case of $(\alpha, AR)=(22^{\circ}, 6)$. The crossflow velocity $u_y$ along $(x,y,z)=(2,-0.3,0-7)$ over time is presented in figure \ref{fig:DMD}(\textit{a}). For $z\gtrsim 5$, which corresponds to the tip vortex region, the velocity remains steady over time. 
For $0\leq z\lesssim5$, the variations in the velocity indicate the occurrence of the periodic vortex shedding. 
Along with the vortical structures in figure \ref{fig:ARAoAEffect} for $(\alpha,AR)=(22^{\circ},6)$, vortex dislocation is clearly observed from the discontinuities in the spanwise vortices.
With the point of dislocation translating repeatedly from $z\approx3.5$ to $z\approx 1$, the wake also changes over time, as observed from the iso-surfaces of $Q=1$ in figure \ref{fig:DMD}(\textit{c}). At $t=95$, the spanwise vortical structures appear intact. Around $t=100$, vortex dislocation starts to develop in the near wake. By $t=105$, the formation of two shedding cells are clearly visible. 

To gain additional insights, we apply Fourier transform to the crossflow velocity $u_y$ at each spanwise section, as shown in figure \ref{fig:DMD}(\textit{b}). We use $\hat{u}_y$ to denote the power spectral density (PSD) of the velocity fluctuation. The two spanwise shedding structures are found to be associated with different frequencies. The structures closer to the mid-span shed with a frequency of $St_1=0.439$ ($St\equiv fc/U_{\infty}$, where $f$ is the dimensional shedding frequency). On the other hand, the structures closer to the wing tip shed with a lower frequency of $St_2=0.372$. To identify the vortical structures associated with these two frequencies, we perform dynamic mode decomposition \citep{schmid2010dynamic,rowley2009spectral} on the $Q$ criteria.  As shown in figure \ref{fig:DMD}(\textit{d}), the modal structures corresponding to $St_1$ is composed of the columnar vortex cores aligned with the spanwise direction. In contrast, the mode with frequency $St_2$ presents three-dimensional modal structures comprised of skewed vortex cores as well as the streamwise vortical structures. These two modes overlap over $1\lesssim z \lesssim 3$. As time evolves, the time-varying phase between these two modes results in the changes in the vortical structures, as observed in figure \ref{fig:DMD}(\textit{c}).

\begin{figure}
\centering
\includegraphics[scale=0.47]{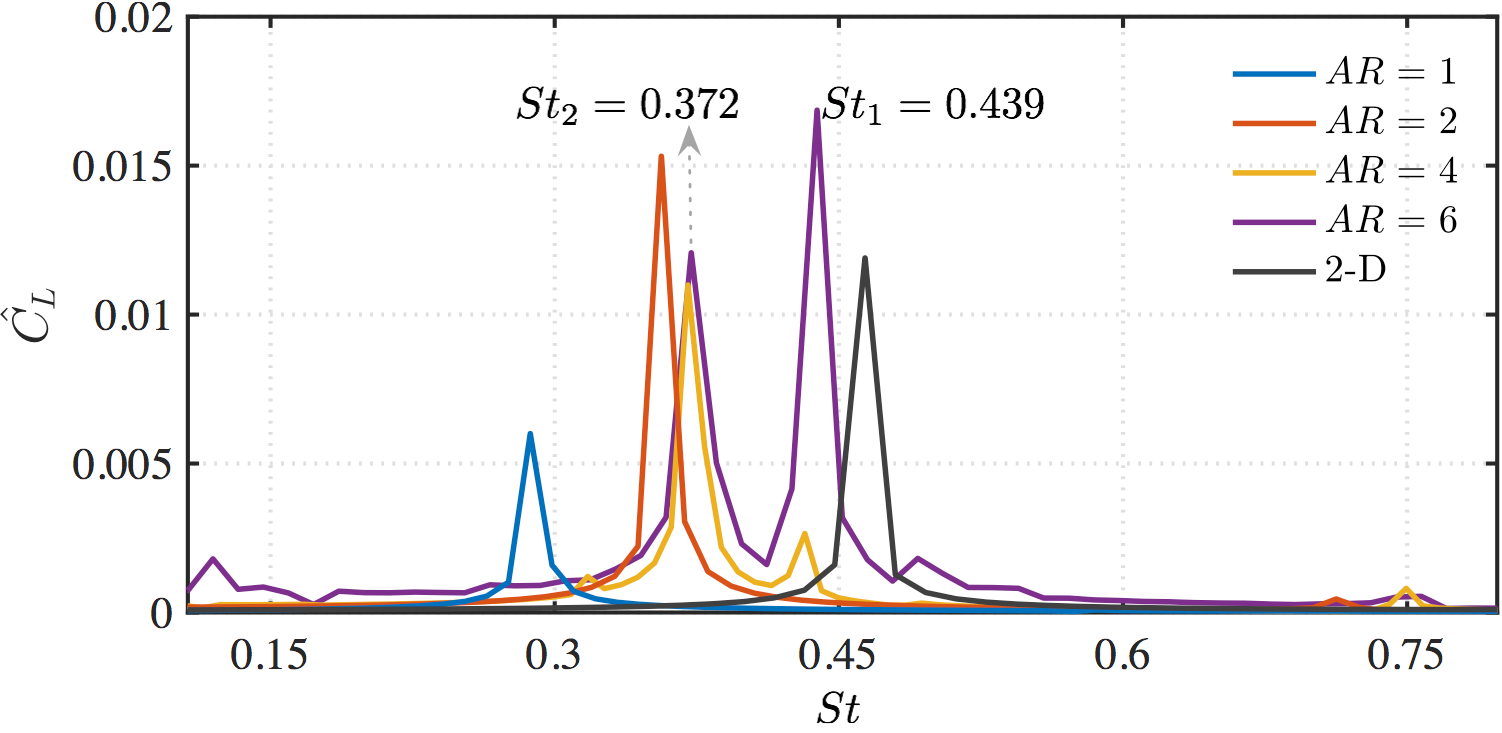}
\caption{Frequency spectra of lift coefficients at $\alpha=22^{\circ}$ for different aspect ratios. $\hat{C}_L$ denotes the PSD of the lift coefficient. The PSD of the two-dimensional case is scaled down by a factor of 10 for visual clarity.}
\label{fig:PSD_CL}
\end{figure}

The dominant frequencies identified in the wake for $(\alpha,AR)=(22^{\circ},6)$ are also detected in the corresponding lift spectrum, as shown in figure \ref{fig:PSD_CL}. Also overlaid are the lift spectra for other aspect ratios with $\alpha=22^{\circ}$. The dominant frequency $St_1$ for $AR=6$ is close to the frequency in the two-dimensional case. This suggests the two-dimensional nature of the corresponding modal structure (mode 1) as shown in figure \ref{fig:DMD}(\textit{d}). 
On the other hand, $St_2$ almost coincides with the primary frequency in the cases of $AR = 2$ and $4$, hinting that the lower shedding frequency is caused by the slow down of shedding from the tip effects. The downward induced velocity from the tip vortex slows the shedding of the neighboring vortical structures.  For the much lower $AR=1$ case, the strong interaction between the tip vortex and the unsteady vortices further slows down the vortex shedding, resulting in a lower shedding frequency.

The lift spectra for different $AR$ reveal how the tip effects influence the wake dynamics along the span. For $AR \lesssim 4$, the tip effects dominate over the whole span of the wing, as discussed in \S \ref{sec:keyFeatures} and \S \ref{sec:AoAAREffect}.
For a higher aspect ratio, the tip effects become weaker over the extended region of $z \gtrsim 4$. This allows nominally two-dimensional structures to develop near the mid-span and compete with the three-dimensional shedding near the tip, giving rise to vortex dislocation.  Based on the insights obtained from spectral analysis, we are able to identify mainly three wake regions along the spanwise direction for large-aspect-ratio wings: (i) the tip vortex region as a direct consequence of the wing tip, (ii) the nominally two-dimensional shedding region near the mid-span (mode 1) where the influence of the tip vortex is weak, and (iii) the three-dimensional shedding region (mode 2) that are influenced by the interaction between regions (i) and (ii).

\subsection{Classification of the wake}
\label{sec:classification}

\begin{figure}
	\centering
	\includegraphics[scale=0.5]{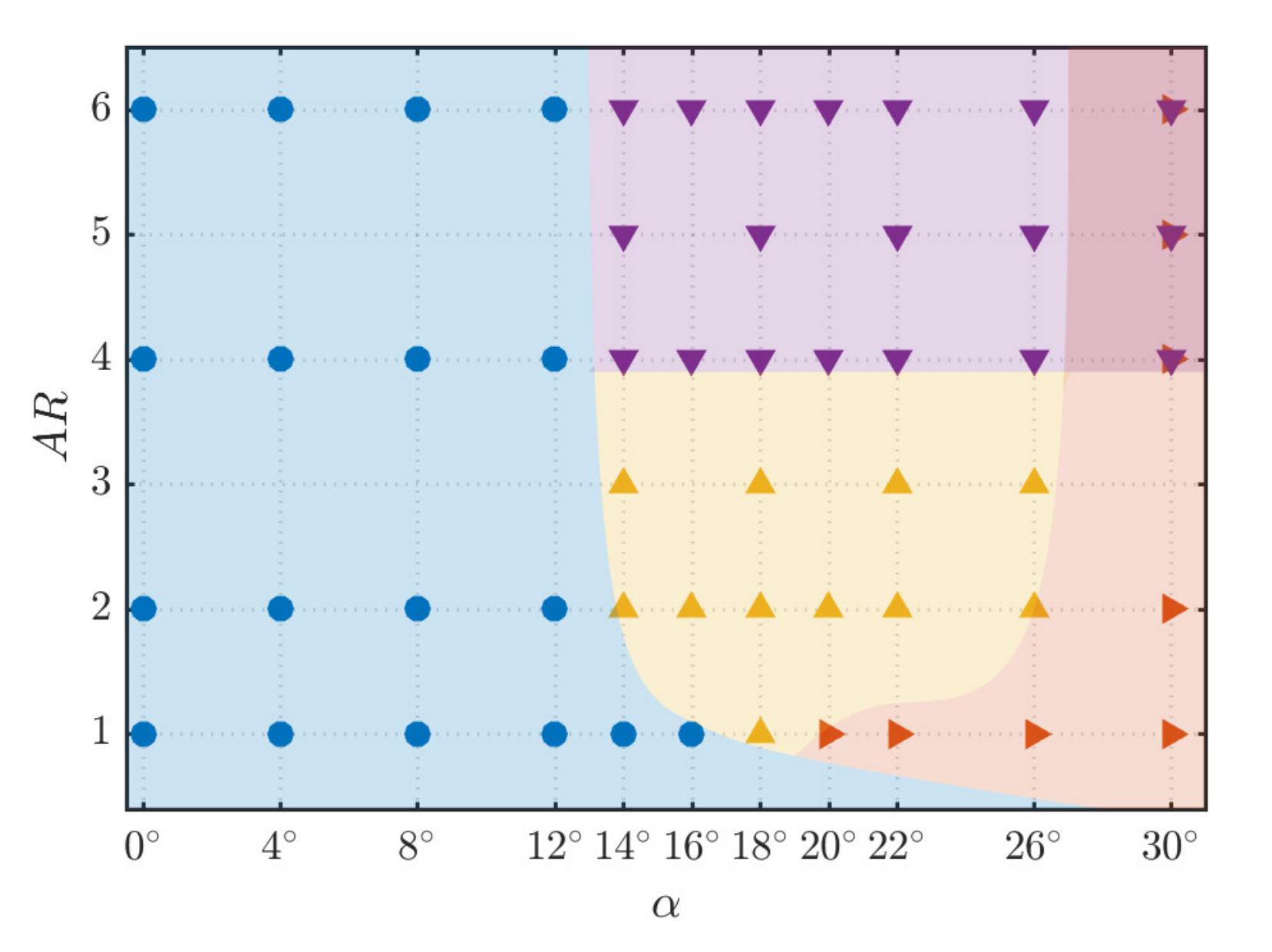}
	\caption{Classification of the wake. \protect\markerone: stable flow; \protect\markertwo: steady tip vortex and unsteady shedding vortices; \protect\markerthree: unsteady tip vortex; \protect\markerfour: vortex dislocation.}
	\label{fig:regime}
\end{figure}

We summarize our wake characterization over aspect ratio and angle of attack in figure \ref{fig:regime}.  At $Re = 400$, the wake remains stable for $\alpha \lesssim 12^{\circ}$ for the aspect ratios considered in this study.  For $AR = 1$, the flow can remain steady for even a higher angle of attack of $\alpha = 16^{\circ}$ due to the stabilizing tip effects.  We shade in blue the conditions for stable wake in figure \ref{fig:regime}.  The transition from steady to unsteady flow is attributed to the Hopf bifurcation associated with the periodically shedding vortical structures \citep{ahuja2007low,taira2009three}.

With the increase in angle of attack, the vortex sheet generated from the leading and trailing edges form shedding vortices, as highlighted by the yellow region in figure \ref{fig:regime}.  
At this stage, the downwash provided by the tip vortex is no longer able to keep these wake vortices stationary.
The wake here is characterized by the presence of two types of vortices: the unsteady shedding vortices and the tip vortex. The unsteady vortices take the form of spanwise vortex cores with their braid-like structures forming closed loops as described in \S \ref{sec:keyFeatures}.

For higher-aspect-ratio wings, the unsteady wake exhibits an additional feature of vortex dislocation.  With the tip vortex imposing downwash on the shedding vortices near the tip, their shedding frequency becomes lower than that of the shedding structures over the mid-span region.  When the discrepancy between these two frequencies become significant, the spanwise shedding processes become desynchronized, resulting in vortex dislocation as shown in figure \ref{fig:DMD}(\textit{b}).  This phenomenon is observed for $AR \gtrsim 4$ and is shaded in purple in figure \ref{fig:regime}.  This type of vortex dislocation is ubiquitous for bluff body flows with spanwise inhomogeneity \citep{eisenlohr1989vortex, williamson1989oblique, noack1991cell, techet1998vortical, lucor2001vortex}. 

As we consider higher angles of attack, the tip vortex exhibits noticeable undulation in the streamwise direction due to its strong interaction with the unsteady shedding vortices.  Such cases are observed over the red region in figure \ref{fig:regime}.  We notice that the transition curve separating the unsteady tip vortex region from the yellow and purple regions is similar to the periodic/aperiodic transition in \citet{taira2009three}. This suggests that the strong interaction between the tip vortices and the unsteady shedding vortices provides a route for the flow to transition to an asymmetric wake, if the symmetry plane is removed.  Such transition is likely caused by the out-of-phase oscillations of the tip vortices at the two ends of the full wing.  Unsteady tip vortex can also interact with the spanwise vortices experiencing vortex dislocation to create a complex wake as indicated by the top right subregion of figure \ref{fig:regime}.  This complex wake generates finer-scale vortical structures from the rich interactions among the tip vortices and the large shedding vortices.

\subsection{Aerodynamic forces}
\label{sec:forces}
\begin{figure}
    \centering
    \includegraphics[scale=0.47]{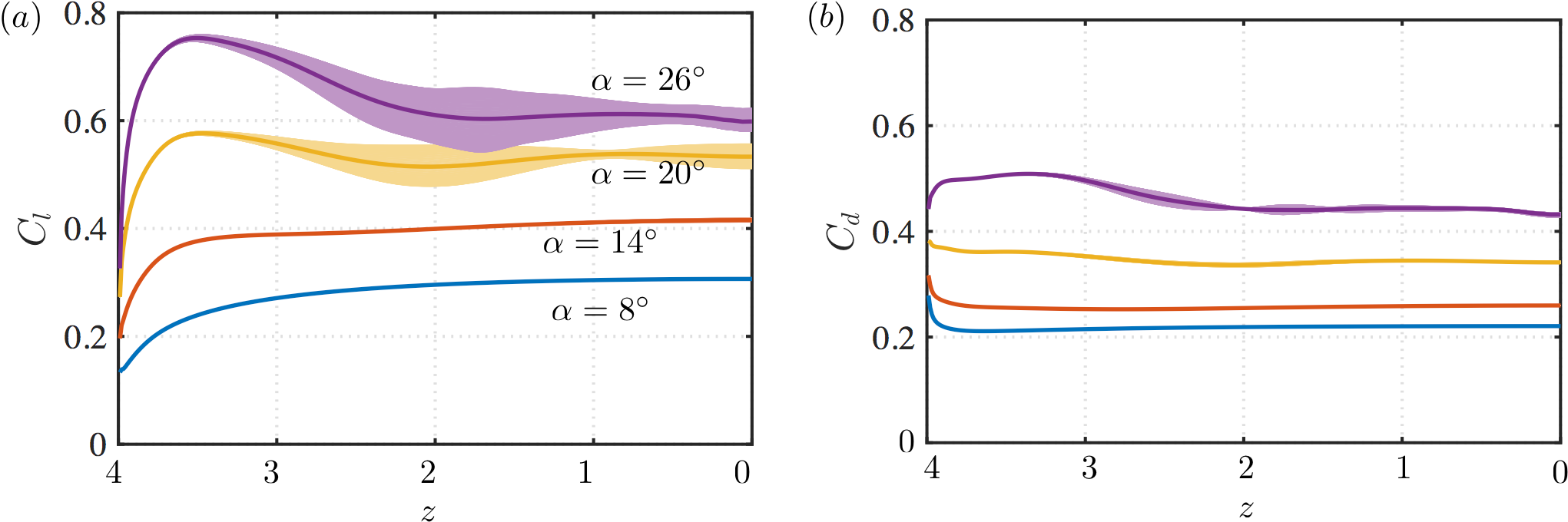}
    \caption{Sectional (\textit{a}) lift coefficients and (\textit{b}) drag coefficients for $AR=4$ with varying angles of attack. The thick curves represent the time-averaged values and the shaded regions represent their temporal fluctuations.}
    \label{fig:sectionalAoAEffect}
\end{figure}

The three-dimensional wake imposes spanwise variations in the sectional distributions of the forces on the wing.  The sectional lift and drag forces for $AR = 4$ are presented for various angles of attack in figure \ref{fig:sectionalAoAEffect}.  For the cases of $\alpha=8^{\circ}$ and $14^{\circ}$, $C_l$ gradually decreases from the mid-span to the wing tip. The drag coefficient remains almost constant over the majority of the span except for the slight increase at the wing tip.

For angles of attack above $20^{\circ}$, considerable temporal fluctuations of $C_l$ appear in the unsteady shedding region, especially near $z\approx 2$. 
The temporal fluctuations in the drag coefficients are significantly lower than the lift coefficients. 
As the flow becomes steady towards the wing tip region, the temporal fluctuation of $C_l$ gradually disappears.  We note here that the time-averaged sectional $\overline{C_l}$ exhibits a peak at $z\approx 3.5$ before it drastically drops at the wing tip. 
This is in contrast to the cases with lower angles of attack, where $C_l$ decays monotonically towards the tip.
This difference in the lift profile will be related to the difference in the near-tip vortical structures in the discussion below.

\begin{figure}
\centering
\includegraphics[scale=0.455]{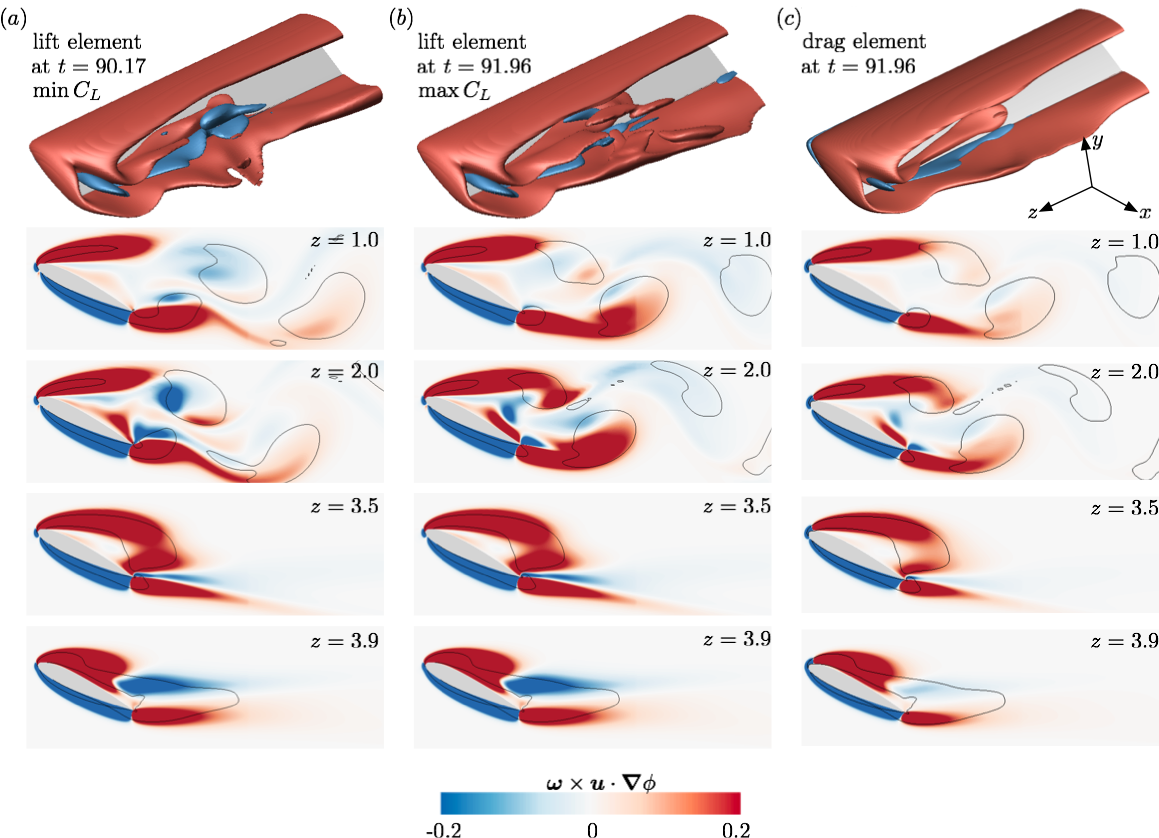}
\caption{
Volume force elements for $(\alpha,AR)=(26^{\circ},4)$. (\textit{a}) Lift elements at $t=90.17$ at which $C_L$ is the lowest. (\textit{b}) Lift force element at $t=91.60$ at which $C_L$ is the highest. (\textit{c}) Drag force element at $t=91.60$. In the top row, the iso-surfaces of lift or drag force elements are shown with the contour level of $\pm 0.2$. The two-dimensional contour plots show the force elements at indicated $z$ planes with the black contours representing the vortex cores ($Q=1$).}
\label{fig:forceElement}
\end{figure}

\begin{figure}
\centering
\includegraphics[scale=0.47]{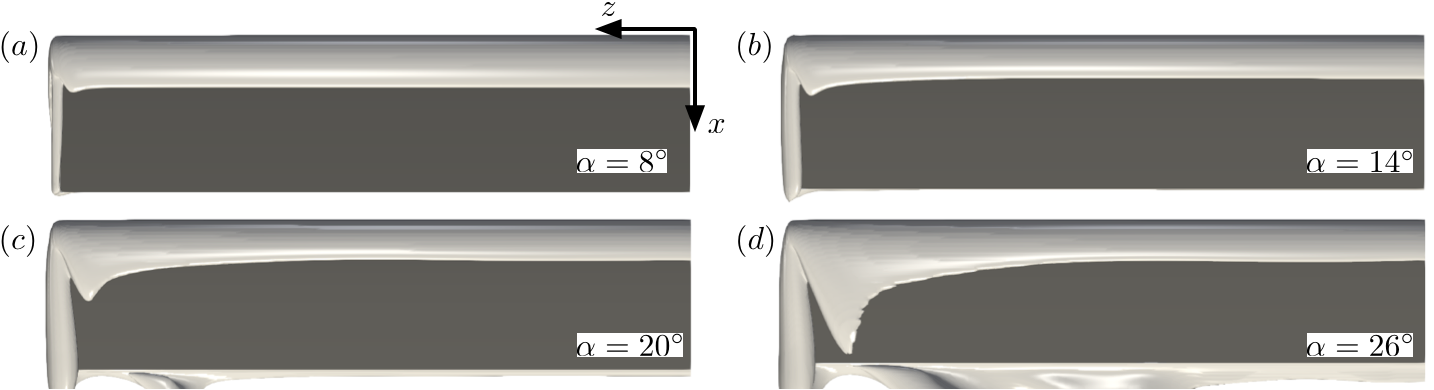}
\caption{
Isosurfaces of $\overline{Q}=3$ for an $AR=4$ wing and varying angles of attack.
}
\label{fig:aveQ3}
\end{figure}

Let us now resort to the force element theory \citep{chang1992potential,lee2012vorticity} to identify the wake structures that exerts aerodynamic forces on the wing. 
In this theory, we utilize an auxiliary potential $\phi_L$ (take the lift force for example) that satisfies the boundary condition 
    $ -\boldsymbol{n}\cdot\boldsymbol{\nabla} \phi_L 
    = \boldsymbol{n}\cdot\boldsymbol{e_y}$. 
By taking the inner product of the Navier-Stokes equation (\ref{equ:NS}$b$) with the potential velocity $\nabla \phi_L$ and integrating over the entire fluid domain $V$, the pressure force term 
    $\int_S p\boldsymbol{n}\cdot\boldsymbol{e_y}\mathrm{d}S$
in equation (\ref{equ:force}) can be expressed as a volume integration of wake variables and a surface integration.  The lift force can then be expressed as
\begin{equation}
    F_L = \int_{V} \boldsymbol{\omega} \times \boldsymbol{u} \cdot \boldsymbol{\boldsymbol{\nabla}}\phi_L \mathrm{d}V 
    + \frac{1}{\Rey}\int_S\boldsymbol{\omega} \times \boldsymbol{n} \cdot (\boldsymbol{\nabla}\phi_L + \boldsymbol{e_y}) \mathrm{d}S.
    \label{equ:liftElement}
\end{equation}
The integrands in the first and second terms on the right hand side are called the volume lift and the surface lift elements, respectively.
The drag force elements can be obtained in a similar fashion by replacing the velocity potential $\phi_L$ with $\phi_D$ and enforce 
$-\boldsymbol{n}\cdot\boldsymbol{\nabla}\phi_D = \boldsymbol{n}\cdot\boldsymbol{e_x}$ 
on the wing surface.  

The lift elements for the case of $(\alpha,AR)=(26^{\circ},4)$ at two time instances, at which the instantaneous lift coefficient is lowest and highest, are shown in figures \ref{fig:forceElement}(\textit{a}) and (\textit{b}), respectively.
In the current study, the contribution of the viscous surface lift element is below 15\% of the total force.  For this reason, we only report below the dominant volume lift element $ \boldsymbol{\omega} \times \boldsymbol{u} \cdot \boldsymbol{\boldsymbol{\nabla}}\phi_L$.
Due to the spatially compact nature of the of the auxiliary potential gradient $\boldsymbol{\nabla}\phi_L$, the lift elements are concentrated in the vicinity of the wing. 
The two free shear layers emanated from both the suction and pressure sides contribute positively to lift, whereas the boundary layer on the pressure surface contributes adversely to the total lift.  Similar observations have also been reported on an impulsively started finite-plate wing by \citet{lee2012vorticity}.  

Comparing the snapshots between two time instances, the difference is observed in the unsteady shedding region. The difference in the lift elements between the two time instances at $z=1$ is subtle. However, at $z=2$, significant amount of negative lift elements appear in figure \ref{fig:forceElement}(\textit{a}) and disappear in figure \ref{fig:forceElement}(\textit{b}). This explains the larger lift fluctuations at $z\approx 2$ than the other regions, as observed in figure \ref{fig:sectionalAoAEffect}(\textit{a}).

Near the tip at $z=3.5$, the local flow is steady and the distributions of lift elements at two time instants are indistinguishable from each other.  Noteworthy here is the large distribution of positive lift element near the wing tip giving rise to increase of sectional lift at around $z = 3.5$ in figure \ref{fig:sectionalAoAEffect}.  Inward from the tip vortex is a vortical structure that forms by rolling up the leading-edge vortex sheet, which is seen from figure \ref{fig:aveQ3} for higher angles of attack. 
This also explains the absence of such local lift increase at lower angles of attack, as the rolled-up vortical structure from the leading edge does not form next to the tip vortex.

Farther towards the tip at $z=3.9$, negative lift elements emerge in the near wake and the positive lift elements diminish, giving rise to the drastic drop in the sectional lift force.  It is noteworthy that the trailing part of the tip vortex barely contributes to the aerodynamic forces. This is not only because of the fast decay of the velocity potential $\boldsymbol{\nabla}\phi_L$, but also due to the fact that in the core of the tip vortex, the velocity vector and the vorticity vector are both aligned with the streamwise direction. The cross-product $\boldsymbol{\omega}\times\boldsymbol{u}$ is small in magnitude, thus making the integrand in equation (\ref{equ:liftElement}) negligible.

The volume drag elements for the same case are shown in figure \ref{fig:forceElement}(\textit{c}). In general, the drag elements share similar patterns with that of the lift elements, where the two separated free shear layers contribute positively to the total drag and the boundary layer on the pressure side contributes negatively. 
The iso-surfaces of drag elements, compared with the lift element at the same time instances, is devoid of the small-scale structures.
Moreover, the intensity of the drag elements is lower compared to the lift elements. Thus, the integrated drag is smaller than lift in both temporal fluctuations as well as time-averaged value, as depicted in figure \ref{fig:sectionalAoAEffect}.

\begin{figure}
    \centering
    \includegraphics[scale=0.47]{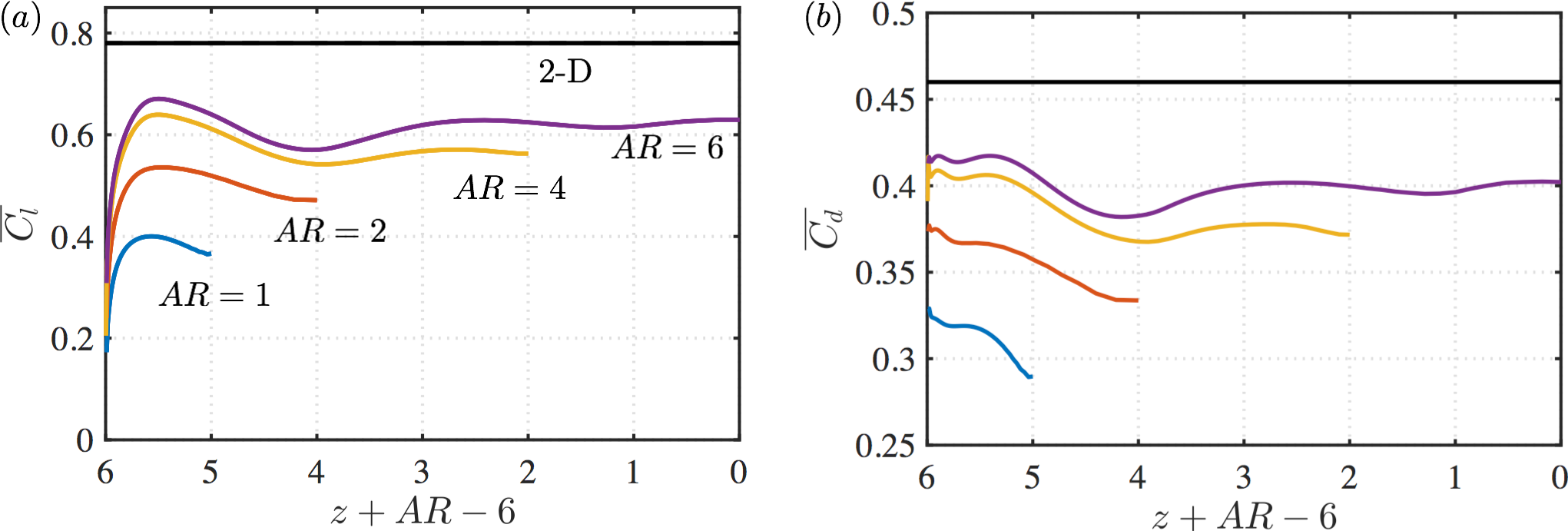}
    \caption{Influence of aspect ratio on the sectional (a) lift and (b) drag coefficients at $\alpha=22^{\circ}$.  The force coefficients are aligned from the wing tip.} 
    \label{fig:SectionalAREffect}
\end{figure}

We further examine the effect of aspect ratio on the sectional lift coefficients for $\alpha=22^{\circ}$ in figure \ref{fig:SectionalAREffect}.  Here, the force coefficients are presented by aligning the distributions from the wing tip.  
With the increase in $AR$, the lift coefficients increase, although they remain below the two-dimensional lift value at the same angle of attack. Focusing in the tip region, the peak in sectional lift near the tip is observed for finite-aspect-ratio wings. This is caused by the presence of the leading-edge vortical structure near the tip as discussed above.  Such structure gives rise to the increase in lift about 0.3 to 0.4 chords away from the tip.
Similar to the lift coefficients, the drag coefficients increase with aspect ratio. However, the spanwise variations for the drag coefficients are relatively small compared to lift. The increase of the sectional $\overline{C_d}$ near the tip is also observed for all aspect ratios at the angle of attack of $\alpha=22^{\circ}$.

\begin{figure}
\centering
\includegraphics[scale=0.46]{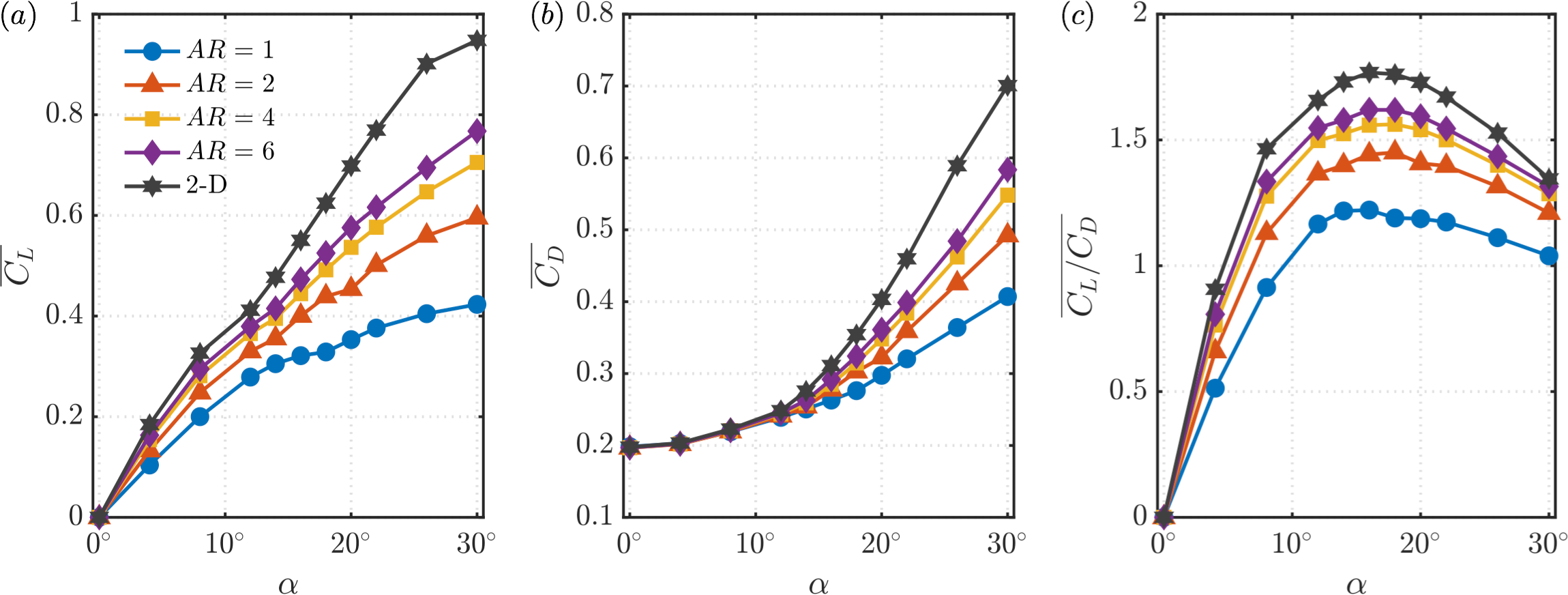}
\caption{
Time-averaged (\textit{a}) lift coefficient, (\textit{b}) drag coefficient, and (\textit{c}) lift-to-drag ratio.}
\label{fig:CLCD}
\end{figure}

At last, let us summarize our aerodynamic force data for the finite-aspect-ratio wings.  The time-averaged lift coefficients $\overline{C_L}$, drag coefficients $\overline{C_D}$, and the lift-to-drag ratios $\overline{C_L/C_D}$ are plotted in figure \ref{fig:CLCD}.  Overall, we notice that the lift and drag coefficients on the wing are smaller for lower-aspect-ratio wings and tend toward the two-dimensional limit for higher-aspect-ratio wings.

For the lift coefficients shown in figure \ref{fig:CLCD}(\textit{a}), we observe that the tip effects reduce lift even at low angles of attack.  For post-stall flows, the influence of the tip vortices on lift becomes more noticeable with wider variations in the lift values with respect to $AR$. At low Reynolds number of 400, lift coefficients monotonically increase for the range of angles of attack considered.  The enhanced lift at higher angles of attack is created by the vortical structures that form above the wing as seen in figure \ref{fig:forceElement} \citep{taira2009three,lee2012vorticity}.  We can observe that above $\alpha \approx 12^\circ$, there is a change in the lift slope which is caused by the emergence of unsteadiness in the wake, as shown in figure \ref{fig:regime}.

Next, let us examine the variation of drag coefficient over the angle of attack.  As in the case with lift, we observe that the drag coefficients become higher as the aspect ratio increases.  This is due to the flow being able to pass around the tip to reduce the drag coefficient for lower-aspect-ratio wings.  However, we notice a striking feature that the drag coefficient is almost independent of the aspect ratio up to $\alpha \approx 12^\circ$.  This suggests that the drag force is primarily due to viscous effects even though the flow is separated at the moderate angles of attack.  Once the wake becomes unsteady beyond $\alpha \approx 12^\circ$, pressure effects contribute significantly to drag, which then is influenced by the aspect ratio and the resulting unsteady wake profiles.  These observations are also in agreement with the findings for finite-aspect-ratio flat-plate wings from \citet{taira2009three}.

At this low Reynolds number, the lift-to-drag ratio is limited to $\mathcal{O}(1)$, as evident from figure \ref{fig:CLCD}(\textit{c}).  
Another unique feature for such low-$Re$ aerodynamics is the high angle of attack at which the peak lift-to-drag ratio is achieved. These ratios are generally in agreement with those for finite-$AR$ flat-plate wings from \citep{taira2009three}, but achieve their peaks at slightly higher angles of attack for the NACA 0015 airfoil.  Such observations are reported in \citet{torres2004low} for higher Reynolds number of $10^5$.  If the finite wings are to be operated in steady condition, it can still achieve a high lift-to-drag ratio for $\alpha \lesssim 12^{\circ}$.  The ratio can be further increased but unsteadiness in the forces from the wake dynamics needs to be considered. 

\section{Conclusion}
\label{sec:conclusion}

We performed direct numerical simulations of the flow over unswept NACA 0015 wings with straight cut tip at a Reynolds number of 400 with the objective of characterizing the tip effects on the three-dimensional separated flows.
The present study considered a half-span wing model with symmetry boundary condition prescribed at the mid-span.  An extensive parametric study was performed for aspect ratios  of $AR=1$ to $6$ and angles of attack of $\alpha=0^{\circ}$ to $30^{\circ}$. 

Based on these simulations, we have made several observations on the emergence of three-dimensional separation and wake patterns.
The inception of three dimensionality is closely tied to the formation of the tip vortex. 
Vorticity is introduced to the flow from the wing surface in a predominantly two-dimensional manner.
As the vortex sheet rolls up around the wing tip to form the tip vortex, three dimensionality is introduced to the flow but is confined to the near-tip region of the boundary layer.  Once the tip vortex reaches its maturity, the stronger downwash influences the roll-up of the leading and trailing-edge vortex sheets, generating three-dimensional vortical structures in the wake and imposing a three-dimensional flow pattern in the vicinity of the wing surface.

The three-dimensional wake is dependent on the aspect ratio and angle of attack of the finite-aspect-ratio wing.
For wings with low $AR$, the strong downwash from the tip vortex inhibits the roll-up of the leading and trailing-edge vortex sheets over the entire span.  For this reason, the wake is able to remain steady for moderate angles of attack. For wings with higher $AR$, the relatively weakened tip effects near the mid-span are no longer able to keep the leading- and trailing-edge vortex sheets from rolling up.  As a result, unsteady shedding takes place away from the tip. 
These shedding vortices are in the forms of spanwise vortex cores and braid-like structures.  Together they close the vortex loops as they shed into the wake.  Further increasing the $AR$ brings out the locally two-dimensional vortical structures at the mid-span. Their competition with the near-tip three-dimensional shedding vortices results in vortex dislocation.  For high angles of attack, the wake becomes complex with interactions among the wake vortices.  With streamwise oscillations appearing at higher angles of attack, the wake may transition to asymmetric flow about the mid-span in the absence of symmetry boundary condition.
These wake features are mapped over a range of aspect ratio and angles of attack.

The three-dimensional wake dynamics has a significant impact on the sectional distribution of aerodynamics forces.  By using the force element analysis, we identified the dominant vortical structures responsible for exerting lift and drag forces on the wing. Noteworthy here are the high sectional force coefficients near the tip for cases at high angles of attack.  Such local increases in force coefficients are attributed to the stationary vortical structure formed inward from the tip vortex.  From the simulations, the aerodynamic force coefficients and the lift-to-drag ratio were also summarized for a wide range of aspect ratios and angles of attack.  Through a comprehensive analysis of the forces and wake dynamics, the relationship between the aspect ratio and the unsteadiness in the aerodynamic characteristics of the wing were revealed.

This study offered a detailed look into the influence of the tip effects on the formation of three-dimensional wake behind finite-aspect-ratio wings.  The insights obtained from this study forms a foundation for further understanding the complex flow dynamics at higher Reynolds number and wake generation by unsteady wing maneuvers.  Moreover, the knowledge gained here can serve as an important reference to active flow control strategies to stabilize or take advantage of vortical forces.

\section*{Acknowledgement}
We acknowledge the US Air Force Office of Scientific Research (Program Managers: Dr. Douglas Smith and Dr. Gregg Abate, Grant number: FA9550-17-1-0222) for funding this project. 

\appendix
\section{Effect of contour level on the characterization of the tip vortex}
\label{sec:contourLevel}

\begin{figure}
\centering
\includegraphics[scale=0.47]{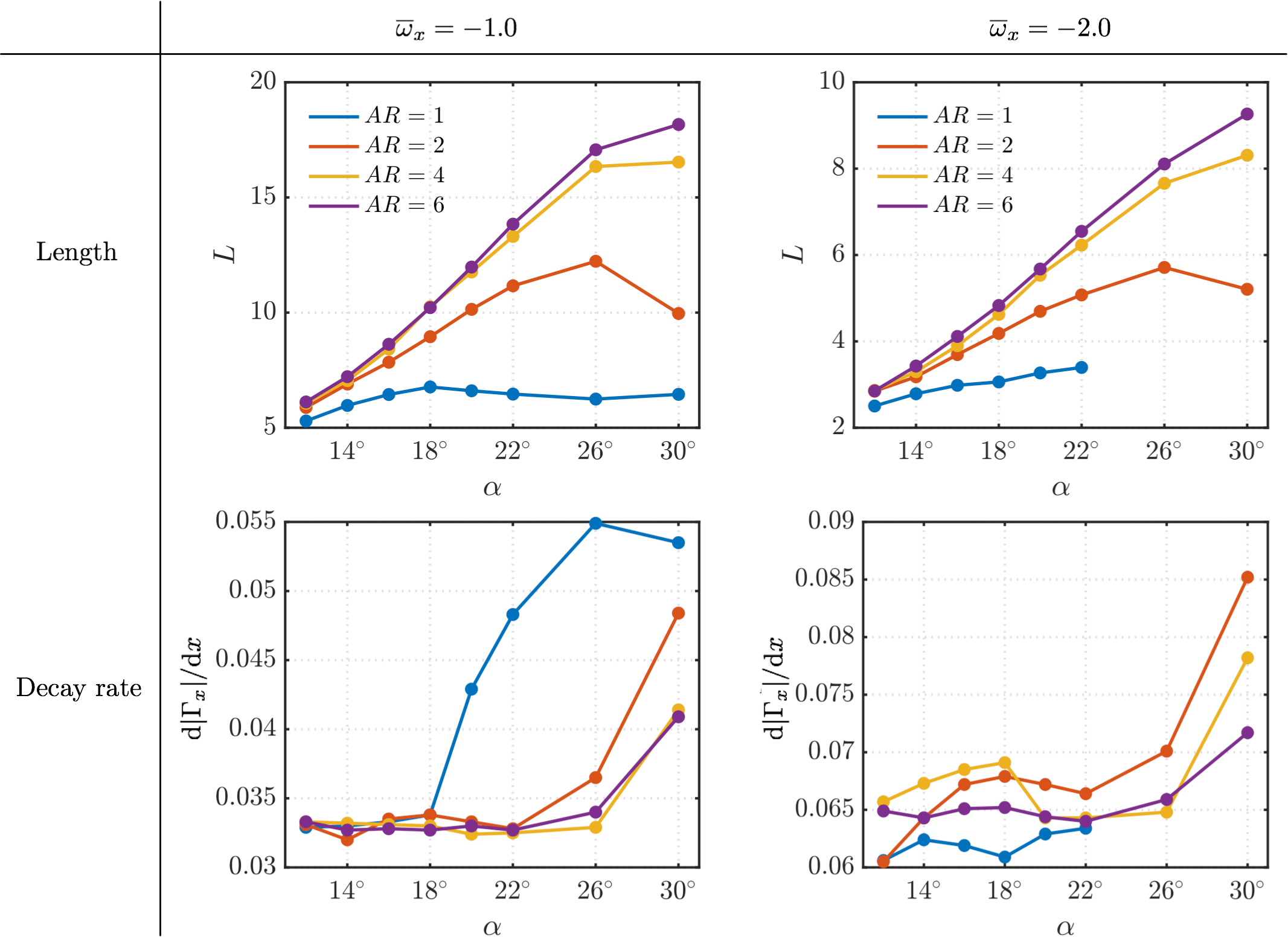}
\caption{Effect of $\overline{\omega}_x$ contour level on the length and decay rate of the tip vortex. }
\label{fig:contourLevel}
\end{figure}

In \S \ref{sec:tipVortex}, the evaluation of the length $L$ and decay rate $\mathrm{d}|\Gamma_x|/\mathrm{d}x$ of the tip vortex involves the selection of $\overline{\omega}_x$ contour level. We have selected $\overline{\omega}_x=-1.5$ as a representative contour level and compiled the characteristics of the tip vortex for various cases. Here, we examine two other contour levels to show that the overall findings are not influenced by the choice of contour level. The lengths and decay rates of the tip vortex based on $\overline{\omega}_x=-1.0$ and $-2.0$ are shown in figure \ref{fig:contourLevel}. As qualitatively observed in figure \ref{fig:tipVortex}(\textit{b}), the absolute values of $L$ and $\mathrm{d}|\Gamma_x|/\mathrm{d}x$ are dependent on the choice of contour level.  The larger the selected value of $|\overline{\omega}_x|$ is, the shorter the length and the larger the decay rate become. However, the variations of $L$ and $\mathrm{d}|\Gamma_x|/\mathrm{d}x$ with respect to $\alpha$, $AR$ and $\Rey$ within each contour level follow the same trends with those based on $\overline{\omega}_x=-1.5$. This suggests that observations we have made in \S \ref{sec:tipVortex} is independent of the choice of the contour level considered in the present study.

\bibliography{reference}
\bibliographystyle{jfm}

\end{document}